\documentclass[authoryear,3p,12pt]{elsarticle}

\usepackage{lscape}

\usepackage{longtable}
\usepackage{amsmath,amssymb}
\usepackage{graphicx}

\usepackage{amssymb}
\usepackage{mathtools}

\usepackage{lipsum}

\usepackage[mathlines]{lineno}

\begin{document}


\begin{frontmatter}

\title{Architectures of planetary systems formed by pebble accretion}

\author[ucla,jpl]{Ryuji Morishima\corref{cor1}}
\ead{morch116@gmail.com}

\cortext[cor1]{Corresponding author}

\address[ucla]{University of California, Los Angeles, Department of Earth, Planetary, and Space Sciences, Los Angeles, CA  90095, USA}
\address[jpl]{Jet Propulsion Laboratory/California Institute of Technology,
Pasadena, CA 91109, USA}

\begin{abstract}
In models of planetary accretion, pebbles form by dust coagulation and rapidly migrate toward the central star.
Planetesimals may continuously form from pebbles over the age of the protoplanetary disk by yet uncertain mechanisms. 
Meanwhile, large planetary embryos grow efficiently by accumulation of leftover 
pebbles that are not incorporated in planetesimals.
Although this process, called pebble accretion, is offering a new promising pathway for formation of giant planets' cores, 
architectures of planetary systems formed through the process remain elusive.
In the present paper, we perform simulations of formation of planetary systems using 
a particle-based hybrid code, to which we implement most of the key physical effects as precisely as possible.
We vary the size of a protoplanetary disk, the turbulent viscosity, the pebble size, the planetesimal formation efficiency, 
and the initial mass distribution of planetesimals. 
Our simulations show that planetesimals first grow by mutual collisions if their initial size is the order of $100$ km or less.
Once planetesimals reach $\sim$ 1000 km in size, they efficiently grow by pebble accretion.
If pebble supply from the outer region continues for a long period of time in a large protoplanetary disk, 
planetary embryos become massive enough to commence runaway gas accretion, resulting in gas giant planets.
Our simulations suggest that planetary systems like ours form from protoplanetary disks with moderately 
high turbulent viscosities.
If the disk turbulent viscosity is low enough, a planet opens up a gap in the gas disk and halts accretion of pebbles 
even before the onset of runway gas accretion. 
Such a disk produces a planetary system with several Neptune-size planets. 



\end{abstract}


\end{frontmatter}


\section{Introduction}
In a classic picture of planetary accretion, it has been generally assumed that 
all dust particles are instantaneously transformed into gravitationally-bound km-size planetesimals and 
that planets primarily grow by accumulation of planetesimals. 
Numerical simulations of planetary growth starting with only planetesimals \citep{Inaba2003b,Kobayashi2011,Chambers2014}
showed that formation of giant planets' cores of $\sim 10 M_{\oplus}$ generally requires 
surface densities of planetesimals 5-10 times larger than that for the Minimum Mass Solar Nebula \citep[MMSN;][]{Hayashi1981}.  

The picture of planetary accretion has been dramatically changed recently. 
In a modern picture, pebbles first form by dust coagulation and then start rapid radial migration toward the central star \citep{Birnstiel2010}.
Planetesimals may continuously form from migrating pebbles over the age of the protoplanetary disk 
by some yet uncertain mechanisms \citep{Chambers2016}. 
If a considerable amount of mass remains in pebbles,
large planetary embryos grow efficiently by accumulation of pebbles, a process so called pebble accretion
\citep{Ormel2010c,Lambrechts2012,Ida2016}.
If a planetary embryo grows massive enough, 
it starts rapid growth by accreting surrounding nebular gas, resulting in a gas giant planet \citep{Pollack1996,Ikoma2000}. 
Pebble accretion is offering a new promising pathway for formation of giant planets' cores
\citep{Lambrechts2014,Bitsch2015,Levison2015a,Chambers2016,Matsumura2017}.

\citet{Chambers2016} studied pebble accretion in viscously evolving disks and 
showed that formation of gas giant planets is efficient in gaseous disks with large initial sizes and low turbulent viscosities.
Among studies of pebble accretion, only \citet{Chambers2016} assumed continuous planetesimal formation from pebbles and 
varied initial planetesimal sizes while all other studies placed 1000 km-size or larger planetary embryos in the first place. 
These 1000 km-size planetary embryos are large enough for efficient growth due to pebble accretion.
If initial planetesimals are not large enough, 
they primarily grow up by mutual merging. 
For formation of gas giant planets, small planetesimals need to 
grow up to $\sim$1000 km in size before pebble supply from the outer part of the protoplanetary disk ceases \citep{Chambers2016}.

In the present study, we also study pebble accretion assuming continuous planetesimal formation from pebbles.
Unlike \citet{Chambers2016},  we adopt a simple gas disk model in which the surface density is 
inversely proportional to the distance from the central star
and decays exponentially with time, rather than solving viscous diffusion.  
Viscous spreading probably occurs during
the early stage of disk evolution, while this mechanism is generally too slow to explain observed gas dissipation timescales 
\citep[see][]{Stepinski1998,Chambers2009}.
This implies that some non-viscous mechanisms, 
such as photoevaporation \citep{Owen2012} or magnetocentrifugal winds \citep{Bai2013,Suzuki2014}, 
predominantly work during middle to late stage of disk evolution.
Therefore, we give the evolution of gas surface density independent from the turbulent viscosity, assuming that 
planet formation proceeds mostly in the middle to late stage of disk evolution.
This assumption will make the dependence of final masses of gas giant planets 
on the turbulent viscosity different from that was found by \citet{Chambers2016}.

The numerical method we employ is the particle-based hybrid code for planet formation developed by \citet{Morishima2015}.
This Lagrangian type code can handle detailed orbital dynamics of any solid bodies, since it 
numerically integrates orbits of all types of solid bodies. 
The code can accurately handle spatial non-uniformity of planetesimals and pebbles, 
mean motion resonances, and planetesimal-driven migration.
However, Lagrangian simulations are computationally intense and we need to limit a simulation domain to outside the snow line ($\sim 3$ AU) 
in order to simulate planetary formation processes over the lifetime of a protoplanetary disk, $\sim$ 3--5 Myr.
Our method is similar to the one employed by \citet{Levison2015a}, who adopted disks with a compact size (30 AU) and high surface densities.
Unlike their simulations, we employ disks with large sizes (100-200 AU) and moderately low surface densities, as
large disks are favorable for formation of giant planets \citep{Chambers2016}.
We also vary the initial planetesimal size and the turbulent viscosity whereas those were fixed in \citet{Levison2015a}.

The present paper is organized as follows.
Section~2 describes modifications made to our particle-based hybrid code for handling pebble accretion. 
In Section~3, we introduce our disk model and explain all effects taken into account in our simulations.
Section~4 shows some simulation examples. 
In Section~5, we discuss the parameter dependence of architectures of planetary systems.
Future work will be discussed in Section~6.
Section~7 summarizes main findings.





\section{Simulation codes}

\subsection{Code overview}
We use a particle-based hybrid code developed by \citet{Morishima2015}.
The code has four classes of particles: full-embryos, sub-embryos, planetesimal-tracers, and pebble-tracers.
Pebble-tracers are newly added for the present study and 
we will describe differences between 
planetesimal-tracers and pebble-tracers in detail below.
Orbits of any types of particles are directly integrated.
The accelerations 
due to planetary embryos' gravity are directly calculated by the $N$-body routine. 
The code can handle a large number of small planetesimals/pebbles 
using the super-particle approximation, in which a large number of small planetesimals/pebbles 
are represented by a small number of tracers.


As planetesimals grow, the number of planetesimals in a tracer decreases. Once the number of planetesimals in a tracer
becomes unity and its mass exceeds the threshold mass $M_{t0}$, 
this particle is promoted to a sub-embryo. If the mass of the sub-embryo exceeds $100 M_{t0}$, 
it is further promoted to a full-embryo.  The acceleration of the sub-embryo due to gravitational interactions with surrounding planetesimals 
is handled by the statistical routine in order to avoid artificially strong accelerations on the sub-embryo while
the acceleration of the full-embryo is always calculated by the $N$-body routine.
The algorithms and various tests for the validation of the code are described in detail in \citet{Morishima2015,Morishima2017}.


\subsection{Pebble-tracers}

We introduce pebble-tracers. 
We conventionally define a pebble-tracer as a tracer of small bodies with St$< 2$, where St 
is the Stokes number (Eq.~(\ref{eq:epscd})). A tracer of bodies with St$\ge 2$ is called as a planetesimal-tracer.
The boundary value, St $=2$,  is based on 
the study of \citet{Ormel2012a}, who showed a change of modes of 
pebble accretion at St$=2$ for a low relative velocity. 

Gravitational and collisional interactions between planetesimal-tracers are handled in the statistical routine using
the Keplerian elements of planetesimal-tracers, 
as described in \citet{Morishima2015}.
This approach is inappropriate for pebbles which are strongly coupled with gas. 
The collisional probability of pebbles with a planetesimal is calculated by a new routine, 
where their relative velocities are used instead of their relative Keplerian elements (Section~3.3 and Appendix.~C).
The damping effect on planetesimals due to collisions with pebbles is also included in this routine.
The gravitational and collisional interactions between pebbles and embryos are 
directly handled in the $N$-body routine.
We ignore any gravitational and collisional interactions between pebble-tracers. 
We also ignore gravitational interactions (stirring and dynamical friction) between pebbles and planetesimals. 

It is technically possible to handle collisional interactions between pebble-tracers, although 
simulations in the present paper do not generally have sufficient numbers of pebble-tracers 
to derive accurate mutual collisional rates and resulting size distributions of pebbles. 
Instead of directly handling mutual collisions, 
we adopt two cases for evolution of sizes of pebbles during their radial migration.
The first one adopts a constant St number (except the pebble size is fixed during close encounters with embryos). 
This resembles the outcome of simulations of dust coagulation without any collisional destruction \citep{Birnstiel2012,Sato2016}.
These simulations show that St for the largest pebbles remains around 0.1-1.
The another case adopts a constant pebble size. This makes St decrease as pebbles migrate inward. 
This approximation roughly mimics prevention of growth of pebbles due to destructive collisions \citep{Birnstiel2012,Chambers2016}.
We do not consider porosity change of pebbles, although some studies indicate its importance \citep{Okuzumi2012,Krijt2015}.



\subsection{Collisions of tracers with sub-embryos}


In our previous studies \citep{Morishima2015,Morishima2017},
collisions of tracers with sub-embryos were handled in the statistical routine,
where we did not let a tracer and a sub-embryo merge in the $N$-body routine even if they mutually overlap.
This treatment is found to be problematic if
tracers with small constituent bodies encounter with sub-embryos in a gaseous disk. 
Small bodies are often gravitationally captured by sub-embryos
and need to be merged with sub-embryos in the $N$-body routine. 
To avoid duplicative counting both in the statistical and $N$-body routines,
we handle collisions between tracers and sub-embryos
only in the $N$-body routine in the present study.

This approach has two drawbacks.
First, after a tracer is promoted to a sub-embyro,
a first impact in the $N$-body routine roughly doubles the mass of the sub-embyro. 
This effect makes an artificial kink around $M_{t0}$ in the mass distribution, 
as discussed in \citet{Levison2012}.
This effect is, however, not essential 
as far as we focus on planets that are much more massive than $M_{t0}$. 
Second, the collisional damping effect on a sub-embryo is not correctly handled.
In the $N$-body routine, a tracer fully feels the gravitational force of a sub-embryo while
the sub-embryo feels only the gravitational force of a single constituent planetesimal
in the tracer. During a close encounter between them, the tracer is highly accelerated 
while the sub-embryo has only a slight velocity change.
If we add the entire momentum of a tracer to the sub-embryo after they merge, 
it causes an artificially large acceleration of the sub-embryo. 
To avoid this effect, we only add the momentum a constituent planetesimal/pebble to that of the sub-embryo.
This treatment unfortunately underestimates the collisional damping effect, although 
dynamical friction due to surrounding planetesimals, handled in the statistical routine, 
is generally much more important.



\section{Simulation setup}
\subsection{Effects of gas on orbital evolution}
We first describe gaseous forces acting on particles over a wide size range.
Through the paper, we use the cylindrical coordinates $(r, \theta, z)$.
We assume that the gas surface density $\Sigma_{\rm gas}$ varies as a function of distance $r$ and time $t$ as 
\begin{equation}
\Sigma_{\rm gas}(r, t) = 
500 \left(\frac{r}{\rm 1 \hspace{0.2em} AU}\right)^{-1} \exp{\left(-\frac{t}{\tau_{\rm gas}}\right)} \hspace{0.3em} 
{\rm g \hspace{0.2em} cm^{-2}}, 
\hspace{0.5em} ({\rm for} \hspace{0.5em}  r \le r_{\rm disk}) \label{eq:sigg}
\end{equation}
where $r_{\rm disk}$ is the disk size and $\tau_{\rm gas}$ is the gas dissipation timescale.
We adopt the most typical value $\tau_{\rm gas}=$ 1 Myr suggested from observations of protoplanetary 
disks around nearby solar-type protostars \citep{Ohsawa2015}.

The temperature profile is given as \citep{Hayashi1981}
\begin{equation}
T = 280 \left(\frac{r}{\rm 1\hspace{0.3em}AU}\right)^{-1/2} {\rm K}. \label{eq:temp}
\end{equation}
The disk viscosity $\nu$ is given by the $\alpha$ model \citep{Shakura1973} as
\begin{equation}
\nu = \alpha c h_{\rm gas}, 
\end{equation}
where $c$ is the isothermal sound velocity and $h_{\rm gas} = c/\Omega$
is the gaseous scale height,  $\Omega = (GM_{\ast}/r^3)^{1/2}$ is the Keplerian frequency, 
$G$ is the gravity constant, and $M_{\ast}$ is the mass of the central star,
which we assume to be the solar mass $M_{\odot}$. 
We adopt the molecular weight of 2.33. This gives $c = 1.0 \times 10^5 (r/1 \hspace{0.2em} {\rm AU})^{-1/4}$ cm s$^{-1}$.
The combination of the radial profiles of $\Sigma_{\rm gas}$ (Eq.~(\ref{eq:sigg})) and $T$ (Eq.~(\ref{eq:temp})) gives a steady state 
(radially constant) mass accretion rate of the global disk toward the central star as 
\begin{equation}
\dot{M}_{{\rm glob}} = 3 \pi \Sigma_{\rm gas} \nu = 
3.7 \times 10^{-9} \left(\frac{\alpha}{10^{-3}}\right)\exp{\left(-\frac{t}{\tau_{\rm gas}}\right)} \hspace{0.3em} M_{\odot} \hspace{0.2em} {\rm yr^{-1}}.
\label{eq:acgl}
\end{equation}

The velocity vectors of  of a solid body and gas are given by 
$ \mbox{\boldmath $v$} $ and 
$\mbox{\boldmath $v$}_{\rm gas}$, respectively. Their relative velocity is 
defined as $\mbox{\boldmath $v$}_{\rm rel} = \mbox{\boldmath $v$} 
- \mbox{\boldmath $v$}_{\rm gas}$.
The aerodynamic drag force per unit mass is given by \citep{Adachi1976}
\begin{equation}
\mbox{\boldmath $f$}_{\rm drag} = 
-\frac{1}{2M}C_{\rm D}\pi s^2 \rho_{\rm gas} 
v_{\rm rel} \mbox{\boldmath $v$}_{\rm rel} = -\frac{\Omega}{\rm St}\mbox{\boldmath $v$}_{\rm rel},
\end{equation}
where $C_{\rm D}$  is a numerical coefficient (Appendix~A), $M$ and $s$ are the mass and the radius of the body,
 $\rho_{\rm gas}$ is the gas density, 
$v_{\rm rel} = |\mbox{\boldmath $v$}_{\rm rel}|$, and St is the Stokes number of the body.
For a small body in the Epstein drag regime, $C_{\rm D}$ and St are given as
\begin{equation}
C_{\rm D,Eps} = \frac{8c_{\rm m}}{3v_{\rm rel}}, \hspace{0.3em} {\rm St_{Eps}} = \frac{\rho_{\rm p} s \Omega}{\rho_{\rm gas}c_{\rm m}},
\label{eq:epscd}
\end{equation}
where $c_{\rm m} = (8/\pi)^{1/2} c$ is the thermal velocity and $\rho_{\rm p}$ is the material density of the solid body.

\begin{figure}
\begin{center}
\includegraphics[width=0.6\textwidth]{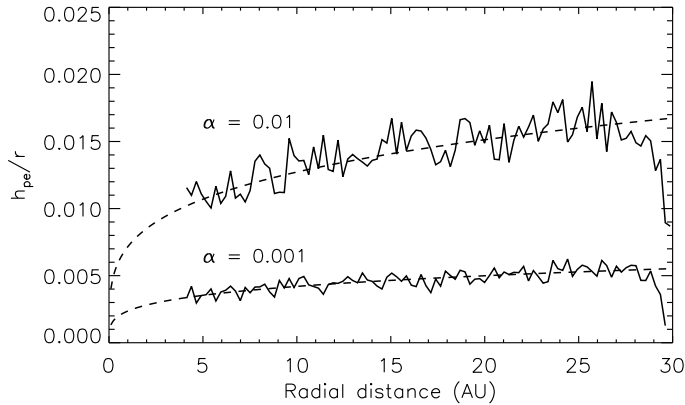}
\caption{Vertical scale height of pebble-tracers for St $= 0.1$ and $\alpha = 0.01$ and 0.001.
Pebble-tracers 
are released at $r=$ 30 AU and deleted at $r=$ 4 AU.
The solid curves are the scale heights from simulations, $h_{\rm pe} = \langle z^2 \rangle ^{1/2}$.
Hundreds of snapshots are superposed to reduce statistical noises.
The dashed curves are the theoretical predictions, 
$h_{\rm pe} = (1+\gamma)^{-1/4}\sqrt{\alpha/(\alpha + {\rm St})}h_{\rm gas}$ \citep{Dubrulle1995}, 
where $\gamma$ is the power-law for the eddy energy spectrum. 
Our simulations show good agreements with theoretical predictions for $\gamma = 3$.
}
\label{fig:hz}
\end{center}
\end{figure}

The gas velocity is given by a sum of the laminar component $\mbox{\boldmath $v$}_{\rm lam} = (0,v_{{\rm lam}, \theta},0)$
and the turbulent component $\mbox{\boldmath $v$}_{\rm tur} = (v_{{\rm tur}, r},v_{{\rm tur}, \theta},v_{{\rm tur}, z})$ as
\begin{equation}
\mbox{\boldmath $v$}_{\rm gas} = \mbox{\boldmath $v$}_{\rm lam} + \mbox{\boldmath $v$}_{\rm tur}, 
\end{equation}
The laminar component $v_{{\rm lam}, \theta}$ in the $\theta$ direction is obtained by the force balance in the $r$ direction \citep{Morishima2010}.
We give the turbulent component using a Lagrangian stochastic model \citep{Wilson1996}.
This model approximately handles the largest turbulent eddies while smaller eddies are ignored.
We assume that turbulence is isotropic, that three velocity components are uncoupled, and 
that the turnover timescale of eddies is $\Omega^{-1}$.
The change of $v_{{\rm tur}, r}$ of gas around each body during the time step $\delta t$ is given as 
\begin{equation}
\delta v_{{\rm tur}, r} = - v_{{\rm tur}, r}\Omega\delta t  + \sigma_{\rm tur}\xi\sqrt{2\Omega \delta t}, 
\end{equation}
where $\sigma_{\rm tur} = (\alpha/3)^{1/2} c$ is the standard deviation of each velocity component (the factor of 3
comes from the three velocity components) and 
$\xi$ is the Gaussian white noise, which has the standard deviation of unity ($\langle \xi^2 \rangle = 1$)
and is uncorrelated in time and space. The other velocity 
components, $v_{{\rm tur}, \theta}$ and $v_{{\rm tur}, z}$ are given in a similar fashion.
This simple approach correctly reproduces the scale height of pebbles (Fig.~\ref{fig:hz}).
Since $\mbox{\boldmath $v$}_{\rm tur} $ is calculated for each particle independently, this model 
does not give correct collision velocities between small pebbles (St $\ll 1$) that are tightly coupled with turbulent motion 
\citep[see][]{Ormel2007}.
This is not a problem as we do not explicitly handle collisions between pebbles (Section~2.2).
 


While pebbles are mainly stirred by 
the aerodynamic coupling with turbulent motion,
planetesimals and larger bodies are more affected 
by gravitational interactions with gas density fluctuations \citep{Laughlin2004,Ida2008,Okuzumi2013}.
We ignore this effect except for one run (Run~16) where we adopt the recipe of \citet{Okuzumi2013} (Appendix~B).


Orbital eccentricities and inclinations of massive bodies are damped by tidal interactions with 
the gaseous disk \citep{Papaloizou2000,Tanaka2004,Capobianco2011}.
We use the formula of \citet{Papaloizou2000} for the tidal damping effects.
In the present paper, we ignore Type I and Type II migration due to tidal interactions with the gaseous disk.
We also ignore the gravitational potential of the global gaseous disk.


\subsection{Formation of pebbles and planetesimals}
We adopt a simple, analytic model for pebble formation.
We assume all the solid component in the protoplanetary disk is all in tiny ($\mu$m-size) dust grains at $t = 0$. 
The dust surface density at $t=0$ is assumed as
\begin{equation}
\Sigma_{\rm dust}(r,t=0) = Z_0\Sigma_{\rm gas}(r, t=0), \label{eq:sigd}
\end{equation}
where $Z_0$ is the metallicity $Z = \Sigma_{\rm dust}/\Sigma_{\rm gas}$ at $t = 0$
and we adopt the solar metallicity of $0.014$ \citep{Lodders2010} in the present paper.

 
We assume that tiny dust particles do not radially move but grow by mutual sticking. 
The growth timescale during which $\mu$m-size dust grains grow to mm-size pebbles is \citep{Lambrechts2014} 
\begin{equation}
\tau_{\rm peb}(r,Z) = \frac{4 \log{(s_{\rm peb}/s_{\rm dust})}}{\sqrt{3}\epsilon_{g,d}Z} \Omega^{-1}, \label{eq:tpeb}
\end{equation}
where $s_{\rm peb}$ and $s_{\rm dust}$ are the radii of dust and pebble particles,
$\epsilon_{g,d}$ is the sticking probability for dust-dust collisions.
We set $\log{(s_{\rm peb}/s_{\rm dust})} = 10$ 
and $\epsilon_{g,d} = 0.1$.   This small value of $\epsilon_{g,d}$ gives 
the dust life time consistent with observations while  dust depletion is too rapid for $\epsilon_{g,d} = 1$ .
The dust abundance is assumed to decay on a timescale of $\tau_{\rm peb}$ due to production of pebbles as
\begin{equation}
\frac{\mathrm{d}\Sigma_{\rm peb}}{\mathrm{d}t}  = - \frac{\mathrm{d}\Sigma_{\rm dust}}{\mathrm{d}t}
= \frac{\Sigma_{\rm dust}}{\tau_{\rm peb}}, \label{eq:dspeb}
\end{equation}
where $\Sigma_{\rm peb}$ is the pebble surface density. 
Using  Eqs.~(\ref{eq:sigg}) and (\ref{eq:sigd})--(\ref{eq:dspeb}),  the dust surface density at $r$ as a function of time $t$ is derived as 
\begin{equation}
\frac{\Sigma_{\rm dust}(t)}{\Sigma_{\rm dust} (t = 0)}  = \left[\frac{\tau_{\rm gas}}{\tau_{\rm peb}(r,Z_0)}\left(\exp{\left(\frac{t}{\tau_{\rm gas}}\right)}-1\right) + 1\right]^{-1}.
\end{equation}
The growth timescale of dust particles is long at large distance while it is shortened by gas depletion as $\tau_{\rm peb} \propto Z^{-1}$.
For $t \ll \tau_{\rm gas}$, $\Sigma_{\rm dust}(t) \propto (t/\tau_{\rm peb}(r,Z_0) + 1)^{-1}$ 
while for $t \gg \tau_{\rm gas}$, $\Sigma_{\rm dust}(t) \propto \exp(-t/\tau_{\rm gas})$
as is the case of gas. 


To handle production of pebble-tracers, we make radial grids. If the total mass of pebbles produced in a grid cell during a time interval
exceeds the tracer mass $M_{t0}$, we make a new pebble-tracer at $r$ with a constituent pebble size determined from the assigned St (Eq.~(\ref{eq:epscd})). 
The time interval is measured from the time 
at which the previous pebble-tracer was produced at the same location $r$ to the present time.
Once pebbles form, they start migrating inward due to gas drag. 
In the present study, pebble-tracers are either (1) converted into planetesimals, 
(2) merged with existing planetesimals or embryos, 
or (3) removed at the snow line $r_{\rm snow}$. 
To save the computational time, we perform orbital integration of pebble-tracers 
only in the region inside the cut-off radius $r_{\rm cut}$ ($r_{\rm snow} < r_{\rm cut} < r_{\rm disk}$).
If a pebble-tracer forms at a certain location $r$ beyond $r_{\rm cut}$ at time $t$,
we introduce it to the simulation at $r_{\rm cut}$ and at $t + \delta t$, where $\delta t$ is a time for the pebble-tracer to migrate from $r$ to $r_{\rm cut}$. 
If this pebble-tracer is converted into a planetesimal-tracer before reaching $r_{\rm cut}$, we delete it.


Given large uncertainties in planetesimal formation mechanisms,  efficiencies, and locations, 
we employ a parameterized approach.
Pebbles are converted into planetesimals on the timescale $\tau_{\rm plan}$ as  \citep{Chambers2016}
\begin{equation}
\frac{\mathrm{d}\Sigma_{\rm plan}}{\mathrm{d}t} = \frac{\Sigma_{\rm peb}}{\tau_{\rm plan}}.
\end{equation}
In our simulations, this simply means that a pebble-tracer is converted into a planetesimal-tracer on the average timescale of  $\tau_{\rm plan}$.
We adopt a following form for $\tau_{\rm plan}$: 
\begin{equation}
\tau_{\rm plan} = \tau_{\rm plan1} \frac{1+{\rm St}^2}{2{\rm St}} \frac{\Omega ({\rm 1\hspace{0.2em}AU})}{\Omega(r)}, \label{eq:tplan}
\end{equation}
where  $\tau_{\rm plan1}$ is the planetesimal formation timescale from pebbles with St $= 1$ at $r = $ 1 AU.
The radial velocity of a pebble is given as
\begin{equation}
v_{\rm r} = \frac{1+{\rm St}^2}{2{\rm St}} \eta r \Omega,
\end{equation}
where $\eta$ is the fractional deviation of 
gas velocity relative to the Keplerian velocity.
With the correction factor $(1+{\rm St}^2)/(2{\rm St})$ in Eq.~(\ref{eq:tplan}), 
$\mathrm{d}\Sigma_{\rm plan}/\mathrm{d}t$ becomes independent of St as $\Sigma_{\rm peb} \propto v_{\rm r}^{-1}$.

We assume the following initial mass-frequency distribution of planetesimals: 
\begin{equation}
\frac{\mathrm{d}M}{\mathrm{d}N} \propto M^{-1.6} \hspace{1em} ({\rm for} \hspace{0.3em} 10^{-3}M_0 \le M \le M_0),
\end{equation}
where $\mathrm{d}N$ is the number of bodies between $M$ and $M + \mathrm{d}M$,
the power-law exponent $-1.6$ from simulations of streaming instability is adopted 
\citep{Johansen2015a,Carrera2015,Simon2016,Schafer2017}.
The default value for the largest initial planetesimal is $M_0 = 10^{21}$ g.
This case can reproduce a bump around $10^{21}$ g 
seen in the mass-frequency distributions of asteroids and Kuiper belt objects
\citep{Morbidelli2009,Johansen2015a}. However, we also vary $M_0$,
since the initial size of planetesimals is still a matter of debate \citep[e.g.][]{Weidenschilling2011}.
The initial velocity dispersion of planetesimals is assumed to be 0.1 
times of the escape velocity of the largest initial planetesimal.
The initial semimajor axis of the planetesimal-tracer is chosen 
so that the its $z$-component of the orbital angular momentum 
is the same as its progenitorial pebble-tracer.


\subsection{Collision rate and outcome}

\begin{figure}
\begin{center}
\includegraphics[width=0.7\textwidth]{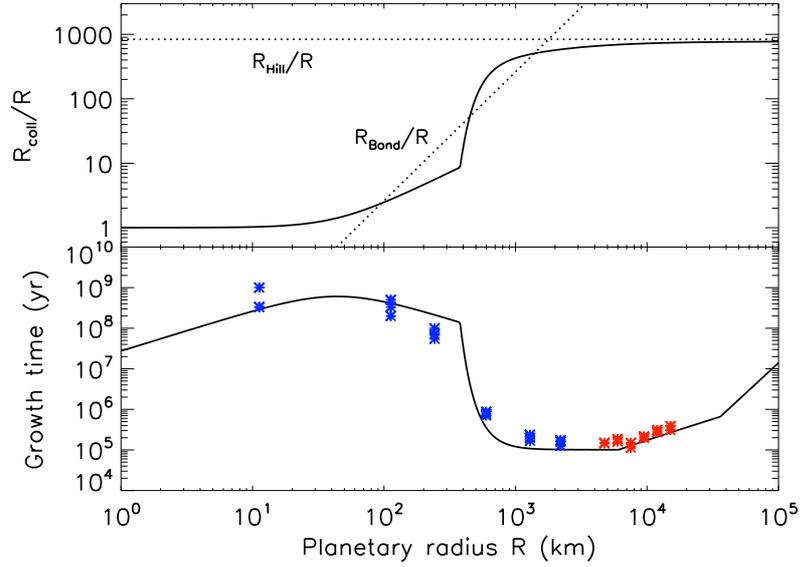}
\caption{Top: Collisional radius $R_{\rm coll}$ for pebble accretion \citep{Ormel2012a} relative to the geometric radius $R$ at $5$ AU for St = 0.1 and $\alpha = 10^{-3}$.
The target body is assumed to have a circular and non-inclined orbit. 
The Hill radius, $R_{\rm Hill}$, and the Bondi radius for pebble accretion, $R_{\rm Bond} = GM/v_{\rm hw}^2$, are shown by dashed lines, 
where $v_{\rm hw} (= 46$ m/s) is the headwind velocity. 
Bottom: The growth time scale $M/\dot{M}$ of an embryo for the pebble mass flux of 100$M_{\oplus}$ Myr$^{-1}$.
The solid line is the theoretical expectation \citep{Ormel2012a}. The asterisks are those from our simulations; the blue 
asterisks are from the statistical routine while the red asterisks are from the $N$-body routine.  
The kink at $R \sim 400$ km is the transition from the hyperbolic regime to the settling regime.
The kink at $R \sim 5000$ km is the transition from 3D to 2D accretion ($R_{\rm coll} = h_{\rm pe}$). 
The kink at $R \sim 40000$ km is caused by a fact that 
a larger embryo sweeps up all pebbles crossing the embryo's orbit ($\dot{M} = $ const). }
\label{fig:f_coll}
\end{center}
\end{figure}

The collision probability between planetesimals (St $\ge 2$) in two different planetesimal-tracers 
is given by the recipe of \citet{Morishima2015}.  
\citet{Ormel2010c}  showed that even for impactors with St $>2$,  the collision probability 
is significantly enhanced due to three body capture,  if the relative velocity is low enough.
We take into account this effect using the prescription 
for the three body regime given by \citet{Ormel2012a}  (Appendix C). 
\citet{Ormel2010c} also showed that
if pebbles (St $< 2$)  encounter with an embryo at low relative velocities, 
they settle toward the embryo at the terminal velocities (the settling regime). 
If the relative velocities of pebbles are large, the gas drag during encounters with the embryo can be ignored (the hyperbolic regime). 
We adopt the collision probabilities of pebbles with planetesimals/embryos 
in the settling and hyperbolic regimes again given by \citet{Ormel2012a} (Appendix C). 

For planetesimal-planeteismal collisions or pebble-planeteismal collisions,
we take into account collisional destruction \citep{Benz1999,Kobayashi2010b}.
The detailed prescription is described in Appendix~D.
The smallest size of collisional fragments is assumed to be the same as the pebble size at each location.

As described in Sections~2, collisions of tracers with full- and sub-embryos are directly 
handled in the $N$-body routine. The routine can automatically reproduce the pebble accretion rate of embryos
without using any analytic recipes.
If a tracer hits an embryo, the outcome is either merging or rebound.
We adopt the boundary velocity for these two outcomes given by \citet{Genda2012}.
If a tracer is gravitationally captured by an embryo during their close encounter, 
we merge them without integrating their orbits until the impact. 
We judge a pair of bodies are gravitationally bound if the mutual distance is less than 0.1 Hill radius and 
the Jacobi integral in Hill units is less than -3.0 \citep[see][for the integral]{Nakazawa1989}. 
We confirmed that these criteria are conservative enough in test simulations. 

The collisional radius of an embryo is enhanced due to 
its dense atmosphere \citep{Inaba2003a,Ormel2012a}.  
The enhanced collisional radius (Appendix~E) is used instead of the radius of the solid core. 
The enhanced collisional radius primarily depends on  
the atmospheric opacity $\kappa$ and the mass accretion rate of the embryo $\dot{M}$.
We adopt $\kappa = $ 0.1 cm$^2$/g.
The accretion rate $\dot{M}$ is derived from the collision log of the embryo.
We search an impact at which the mass of the embryo is closest to 90\% of the present mass.
Using the time $t_{90}$ and the mass $M_{90}$ of the embryo at this impact,
we approximately give the mass accretion rate as
\begin{equation}
\frac{dM}{dt} = \frac{M - (M_{90} +0.5 \delta M)}{t-t_{90}},
\end{equation}
where $\delta M$ is the mass gain by the impact. This mass accretion rate is also used for judging 
the onset of runaway gas accretion to the embryo (Section~3.4). 

Figure~\ref{fig:f_coll} shows the collisional radius and the growth timescale of a target body for pebble accretion (St $=0.1$).
The collisional radius is almost similar to the geometric radius for small planetesimals while 
it significantly increases with radius in the radius range of 100 km to 1000 km.
The growth timescales of target bodies in our test simulations show good agreements 
with the theoretical predictions for all the range of the target radius.


\subsection{Runaway gas accretion and gap opening}

If the embryo's mass exceeds the critical mass, it starts runaway gas accretion
from the global gas disk. The critical mass is given by
\citep{Ikoma2000} as
\begin{equation}
M_{\rm crit} = 
7\left(\frac{\dot{M}}{1\times 10^{-7} M_{\oplus}\hspace{0.2em}{\rm yr}^{-1}}\right)^{0.25}
\left(\frac{\kappa}{1 \hspace{0.2em}{\rm cm}^2 \hspace{0.2em}{\rm g}^{-1}}\right)^{0.25} M_{\oplus},
\end{equation}
where $\kappa$ is the atmospheric opacity.
We ignore the atmospheric mass until the embryo starts runway gas accretion for simplicity, although
accurate atmosphere models \citep{Pollack1996,Ikoma2000} 
show that the atmospheric mass is comparable to the core mass  at the onset of runway gas accretion.

When the embryo mass is relatively small, 
the gas accretion rate for the embryo is regulated by the cooling efficiency of its atmosphere:
\begin{equation}
\dot{M}_{\rm KH} = \frac{M}{\tau_{\rm KH}},
\end{equation}
where $\tau_{\rm KH}$ is 
the Kelvin-Helmholz (cooling) time given as \citep{Ikoma2000}
\begin{equation}
\tau_{\rm KH} = 
10^8 \left(\frac{M_{\rm core}}{M_{\oplus}}\right)^{-2.5}
\left(\frac{\kappa}{1\hspace{0.2em} {\rm cm}^2 \hspace{0.2em}{\rm g}^{-1}}\right) \hspace{0.5em}{\rm yr}.
\end{equation}
When an embryo becomes massive enough, 
the gas accretion rate for the embryo is regulated by the global accretion rate $\dot{M}_{\rm glob}$ (Eq.~(\ref{eq:acgl})).
Numerical simulations of \citet{Lubow2006} showed that the growth rate of a gas capturing embryo is about 
75--90 \%  of the global disk accretion rate outside its orbit. We adopt 90\% for this ratio and
therefore give the gas accretion rate as
\begin{equation}
\dot{M} = {\rm MIN}(\dot{M}_{\rm KH}, 0.9\dot{M}_{\rm glob}).
\end{equation}
Accordingly, the global accretion rate inside the orbit of 
the gas capturing embryo is subtracted by $\dot{M}$. This correction is important since 
it is found to be common in our simulations that multiple embryos undergo runaway gas accretion in the same time.


A massive embryo opens up a gap around its orbit in the global gaseous disk  \citep{Lubow2006,Duffell2015,Kanagawa2017}. 
Based on angular momentum conservation,
\citet{Duffell2015} derived an analytic formula of the radial profile 
of a gaseous gap induced by an embedded embryo with the semimajor axis $a$ as
\begin{equation}
\Sigma_{\rm gas}(r, K) = \Sigma_{\rm gas}(r, K = 0)\left(1- \frac{0.048f(r)K}{1+0.048K}\sqrt{a/r} \right),
\label{eq:gappro}
\end{equation}
where
\begin{equation}
K = \left(\frac{M}{M_{\ast}}\right)^2 \left(\frac{h_{\rm gas}}{a}\right)^{-5} \alpha^{-1}.
\end{equation}
The function $f(r)$ is the non-dimensional angular momentum flux due to 
shocking of planetary wakes. 
The function $f(r)$ can be described in terms of the scaled distance $d(r)$ from the planet as
\begin{equation}
f(r) =
\left\{ \begin{array}{ll} 
1
& \mbox{(for $d(r) \le d_{\rm sh}$)}, \\ 
\sqrt{d_{\rm sh}/d(r)}
& \mbox{(for $d(r) > d_{\rm sh} $)},\\
\end{array}\right.
\label{eq:fgap}
\end{equation}
where the scaled shock position $d_{\rm sh}$ was calculated by \citet{Goodman2001} as 
\begin{equation}
d_{\rm sh} = 1.89 + 0.53\left(\frac{M}{M_{\ast}}\right)^{-1}\left(\frac{h_{\rm gas}}{a}\right)^{3}. 
\end{equation}
The scaled distance $d(r)$  is given as 
\begin{equation}
d(r)= 0.93\left(\frac{|r-a|}{h_{\rm gas}}\right)^{5/2}. 
\end{equation}
The surface density profiles around orbits of embyos more massive than $2M_{\oplus}$
are modified using Eq.~(\ref{eq:gappro}). 
If multiple embryos open gaps, we calculate the surface densities reduced by gap opening embryos individually 
and takes the lowest value at each radial location. 
Note also that \citet{Duffell2015} ignored the effect of gas accretion to the embryo for derivation of Eq.~(\ref{eq:gappro}).
Thus, the surface density profile inside the orbit of a gas capturing embryo is not fully consistent with the global gas accretion rate in our model, 
although this effect is minor for overall growth of embryos.

\begin{figure}
\begin{center}
\includegraphics[width=0.6\textwidth]{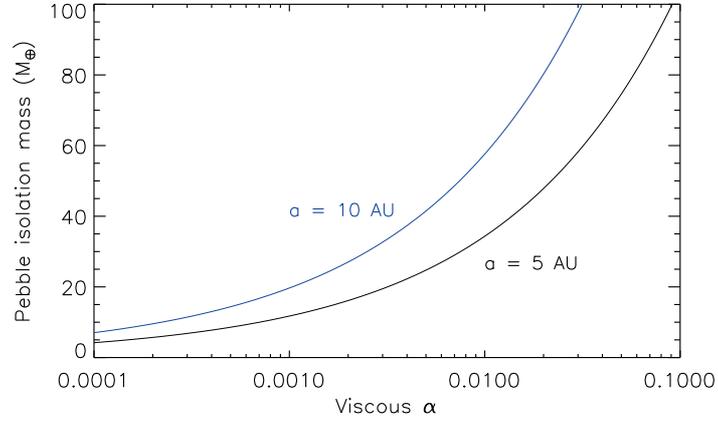}
\caption{Pebble isolation mass. The analytic prescription of  \citet{Duffell2015}  for gap radial profiles is adopted. }
\label{fig:mgap}
\end{center}
\end{figure}

If the embryo is more massive than a certain threshold mass, 
the rotation velocity of gas exceeds the local Keplerian velocity near the outer edge of the gap.
The super Kepelerian rotation prevents pebbles from inward migration and accretion to the embryo.
Figure~\ref{fig:mgap} shows the threshold mass, called the pebble isolation mass, numerically derived from Eq.~(\ref{eq:gappro}).
\citet{Lambrechts2014b} performed hydrodynamic simulations and showed that the pebble isolation mass is about 
$20M_{\oplus}$ for $\alpha = 6 \times 10^{-3}$ and at $a =$ 5 AU.
This is roughly consistent with the value in Fig.~3, $M \simeq 27 M_{\oplus}$.
The pebble isolation mass decreases with decreasing the viscosity. 
For example,  it is about  $12 M_{\oplus}$ for $\alpha = 1 \times 10^{-3}$ at $a =$ 5 AU. 
If the viscosity is further lower, embryos can reach the pebble isolation mass even before they start runaway gas accretion. 

\begin{table}
\begin{center}
\footnotesize
\begin{tabular}{|cccccccc|} \hline
Run~        &  $\alpha$                & St      & $M_{0}$        & $\tau_{\rm plan1}$ & $r_{\rm disk}$ & $r_{\rm snow}$ & Turb. torque  \\ 
                &                                &          &  (g)                &  (yr)                        &  (AU)               &(AU)                  &                       \\ \hline
1             & $3 \times 10^{-3}$   & 0.1   &   $10^{21}$   & 100                       & 200                 & 3                       & No                 \\
1b            &                               &           &                      &                               &                         &                          &                        \\
1c            &                               &            &                     &                               &                         &                          &                        \\
2              &  $1 \times 10^{-4}$ &           &                     &                               &                        &                          &                        \\
3              &  $3 \times 10^{-4}$ &           &                     &                               &                        &                          &                        \\
4              &  $1 \times 10^{-3}$ &           &                     &                               &                        &                          &                        \\
5              &  $1 \times 10^{-2}$ &           &                     &                               &                        &                          &                        \\
6             &                                 &           &   $10^{13}$  &                               &                        &                          &                        \\
7             &                                 &           &   $10^{17}$  &                               &                        &                          &                        \\
8             &                                 &           &   $10^{25}$  &                               &                        &                          &                        \\
9             &                                 &           &                     &   1000                    &                        &                          &                        \\
10           &                                 &            &   $10^{25}$ &   1000                    &                        &                          &                        \\ 
11           &                                 &  0.03   &                     &                               &                        &                          &                        \\ 
12           &                                 &  0.3    &                      &                               &                        &                          &                        \\ 
13           &                                 &  0.03(a/{\rm 3\hspace{0.2em}AU}) &  &     &                        &                          &                        \\ 
14           &                                 &          &                      &                               &  100                  &                          &                        \\ 
15           &                                 &          &                      &                               &                          &    4                    &                        \\ 
16           &                                 &          &                      &                               &                          &                         &     Yes                   \\ 
\hline					                                                               
\end{tabular}
\end{center}
\caption{Input parameters. The values at empty spaces are the same as those for Run~1. For Run~13, St $=0.2$ for $a > r_{\rm cut} = 20$ AU.}
\label{tb:param}
\end{table}

\section{Simulation runs}

The input parameters for 18 simulations we performed are summarized in Table~1.
Run~1 is a base line case and we vary one or two different parameters from the base-line case for other runs.
For $\tau_{\rm plan1} = 100$ yr, roughly a half of pebbles are converted into planetesimals 
before reaching the snow line $r_{\rm snow}$, provided that  they neither merge with existing planetesimals/embryos nor are trapped at the edges of planetary gaps. 
Planetesimals and pebbles are removed if their semimajor axes become less than $r_{\rm snow}$. 
We do not remove embryos even inside the snow line, as they are likely to gravitationally retain
atmospheres consisting of water vapor \citep{Machida2010}. We remove embryos if they are inside 2 AU.
The time step of orbital integration is 7.5 days. The initial mass of a pebble-tracer at its introduction 
is $M_{t0} = 0.02 M_{\oplus}$. This is also the minimum mass of a sub-embryo. 
We performed each run up to $> 5$ Myr.
For Run~2,  we exceptionally stopped the simulation before 4 Myr, as a chain of planets form beyond $r_{\rm cut} (= $ 20 AU) by that time.
Each run takes 2-3 CPU weeks. 
In the following, we first describe processes of planetary growth in detail for three distinctive cases: Runs 1, 2, and 10.

\subsection{Run~1}

\begin{figure}
\begin{center}
\includegraphics[width=0.8\textwidth]{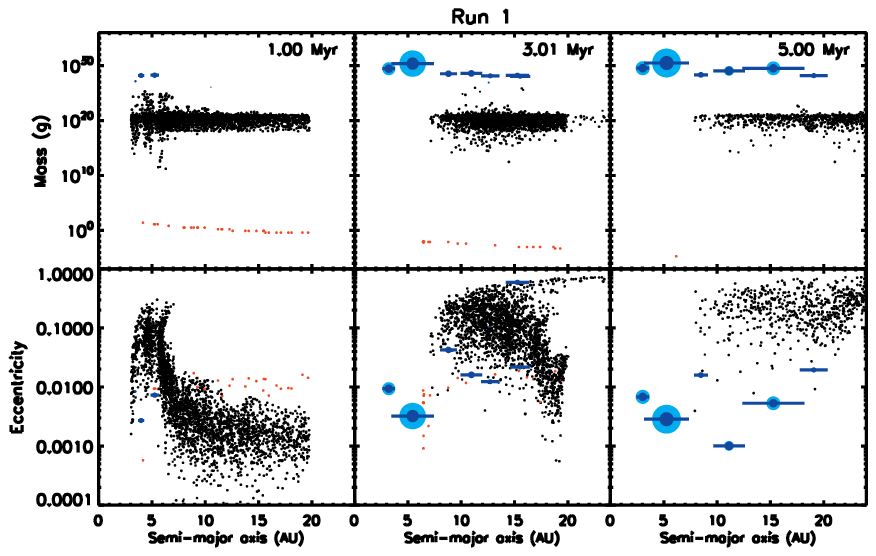}
\includegraphics[width=0.6\textwidth]{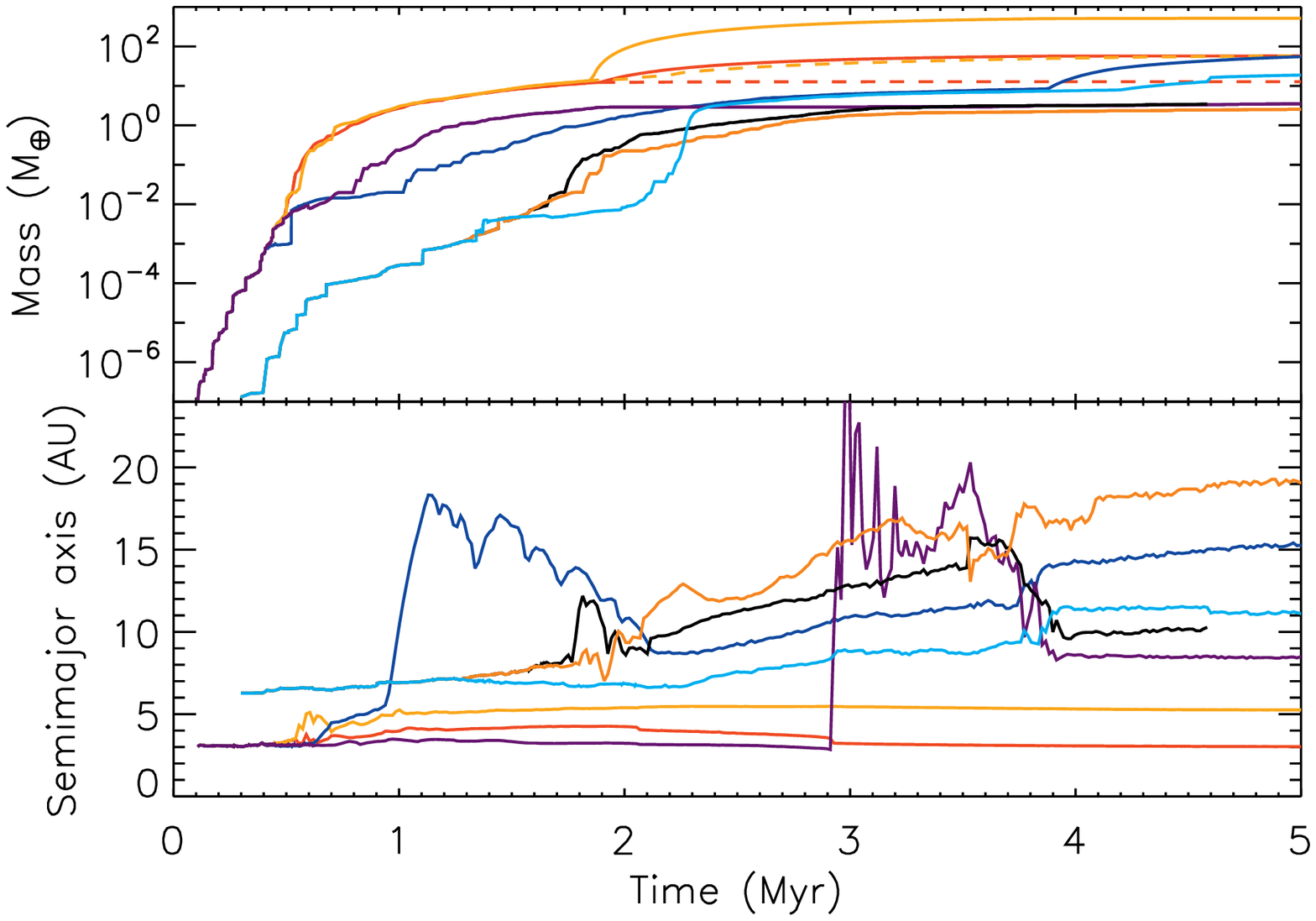}
\caption{(Top) Distributions of mass and orbital eccentricity for Run~1 at three different times. 
Blue, black, and red circles show embryos, planetesimals, and pebbles.
For embryos, the radius of each circle is proportional to the embryo's radius.
For gas capturing embryos,  
each atmosphere is shown by the light-blue annulus around the blue-coded core. 
A half length of a horizontal bar is 10 Hill radii for each embryo with a mass more than $M_{\oplus}$.
Since the orbital eccentricity of each body is derived assuming a Keplerian orbit,  
pebble-tracers have apparent orbital eccentricities of $\sim \eta$.
(Bottom) Time evolutions of masses and semimajor axes of most massive embryos.
Two dotted curves show core masses of two gas capturing embryos.
 }
\label{fig:run1}
\end{center}
\end{figure}

We begin with the base-line case, Run~1.
This is one of a few runs which produced a planetary system similar to our Solar System.
The diameter of the initially largest planetesimal is about 100 km.   The viscosity parameter is moderately high ($\alpha = 3\times 10^{-3}$) .
The distributions of mass and orbital eccentricities at three different times are shown in the upper panel of Fig.~\ref{fig:run1}.
The time evolutions of masses and semimajor axes of massive embryos ($> 2 M_{\oplus}$) are shown in the lower panel 
of Fig.~\ref{fig:run1}. 

Since planetesimals immediately after their formation are not large enough for efficient pebble accretion (see Fig.~\ref{fig:f_coll}),
they first grow by mutual collisions.
Once the largest planetesimals reach $\sim$ 1000 km in size ($M \sim 10^{24}$ g),
they grow more efficiently by pebble accretion than by planetesimal accretion. 
The two most massive embryos reach $1 M_{\oplus}$ in mass around 0.6 Myr near the snow line.
They further grow and reach 
$10 M_{\oplus}$ in mass around 2 Myr, at which they start runway gas accretion.
The outer embryo grows much more rapidly than the inner one,
since the outer one reduces the global gas accretion rate inside its orbit.
The mass of the largest embryo is about 1.3 and 1.5 times of the Jupiter's mass at 3 and 5 Myr, respectively.
The largest embryo opens up a gap in a gas disk and 
accretion of pebbles to the embryo no longer occurs (for $M>60M_{\oplus}$; see Fig.~\ref{fig:mgap}).
However, core growth of the largest embryo occurs even after the onset of runaway 
gas accretion since many planetesimals are scattered toward the largest embryo by outer embryos, which
are not massive enough to efficiently eject planetesimals from the system.

During gas accretion of the inner two embryos, five massive embryos form outside the orbit of the most massive embryo. 
These outer embryos migrate outward, as they grow due to mutual orbital repulsion. 
Some of them also experience rapid planetesimal-driven migration. 
One of the outer embryos has a very large orbital eccentricity due to gravitational perturbations of other embryos 
and is ejected from the system at 4.6 Myr. 
Two of the outer embryos at $\sim$ 11 and $\sim$  15 AU start runaway gas accretion around 4 Myr, and
the outer one reaches $\sim 60 M_{\oplus}$ at 5 Myr.
If the gaseous disk completely dissipates around 4 Myr,  and if outer four embryos merge into two embryos,
they may result in like Uranus and Neptune.
At the end of simulation, the mass-frequency distribution of remaining planetesimals (mostly at $>$ 10 AU) 
is almost the same as the original distribution.

\subsection{Run~2}

\begin{figure}
\begin{center}
\includegraphics[width=0.8\textwidth]{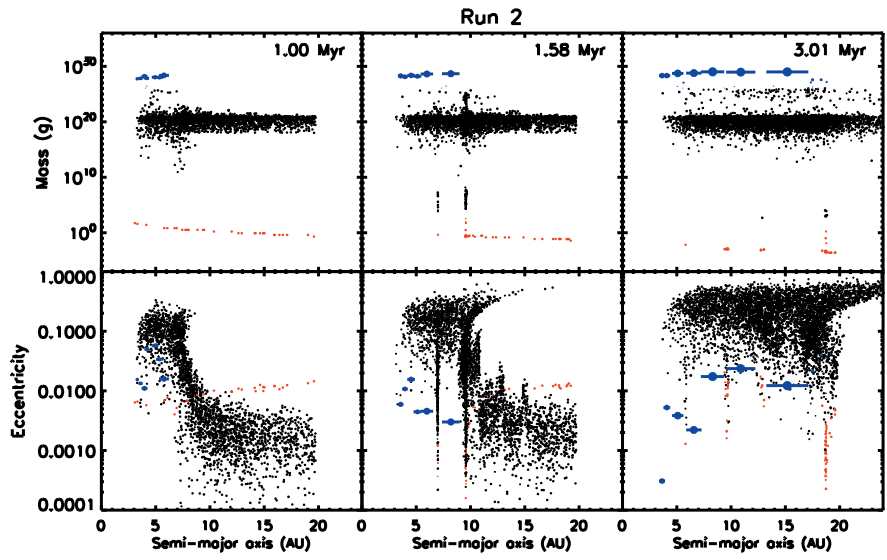}
\includegraphics[width=0.6\textwidth]{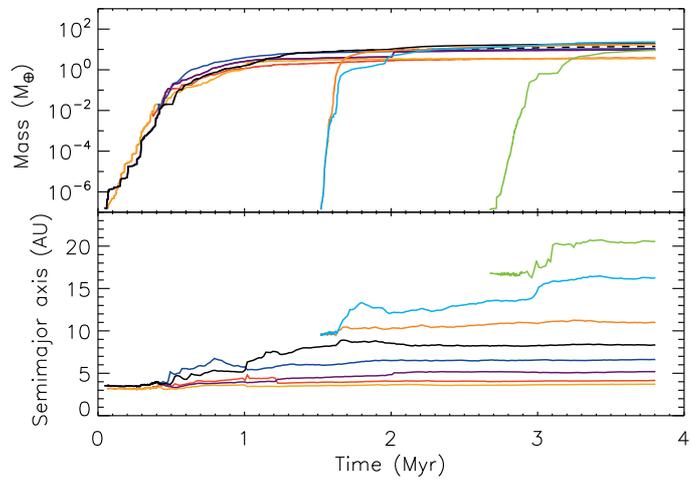}
\caption{The same as Fig.~\ref{fig:run1} but for Run~2.}
\label{fig:run2}
\end{center}
\end{figure}

The input parameter for Run~2 are the same as those for Run~1 except 
that the viscosity parameter for Run~2 ($\alpha = 1 \times 10^{-4}$) is much lower than that in Run~1. 
Figure~\ref{fig:run2} shows the outcome of Run~2.  
Growth of planetesimals during the earliest stage proceeds in a similar fashion to Run~1, 
as we ignore turbulent torques on planetesimals in both Runs~1 and 2. 
Once the largest planetesimals reach $\sim$ 1000 km in size, pebble accretion starts to become efficient. 
1000 km-size planetesimals grow faster than Run~1 on average since the scale height of pebbles is smaller in Run~2 than that in Run~1.
However, in Run~2, growth of massive embryos by pebble accretion is halted due to their gap opening even before they start runaway gas accretion.
Pebbles accumulate at outer gap edges and formation of another massive embryos at the gap edges occurs quickly.
Some of embryos start runway gas accretion, but their growth rates are low due to the low global gas accretion rate. 
As a result of these processes, several embryos with $\sim 10 M_{\oplus}$ form in 3 Myr. Unlike Run1, no gas giant planet forms.

The other important difference between Run~1 and Run~2 is that 
Run~2 has a large number of 1000 km-size planetesimals in the end while almost no 1000 km-size planetesimals exist in the end of Run~1.
In Run~2, these large planetesimals form at the gap edges opened by massive embryos.
Since the gap width normalized by the Hill radius of a gap-opening embryo is much larger in Run~2 than Run~1,
orbits of planetesimals at gap edges are dynamically stable in Run~2. The relative velocities between planetesimals and pebbles
are low at  the gap edge, as both gas and pebbles rotate nearly at the local Keplerian velocity.
Because of these reasons, even 100 km-size planetesimals can efficiently grow to 1000 km in size at the gap edges in Run~2.
In Run~1, in contrast, 1000 km-size planetesimals form only near the snow line in the early stage.






\subsection{Run~10}

\begin{figure}
\begin{center}
\includegraphics[width=0.8\textwidth]{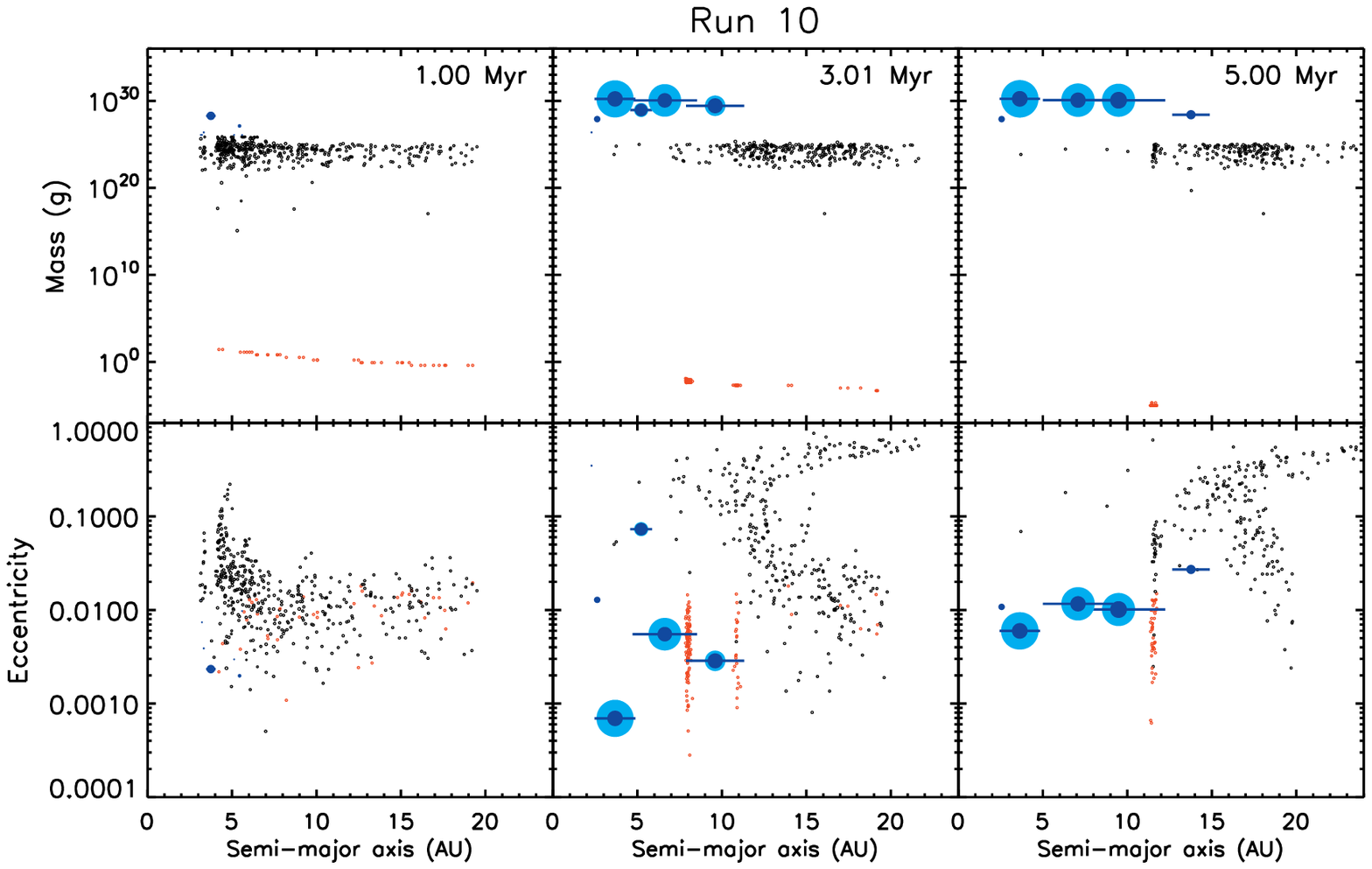}
\includegraphics[width=0.6\textwidth]{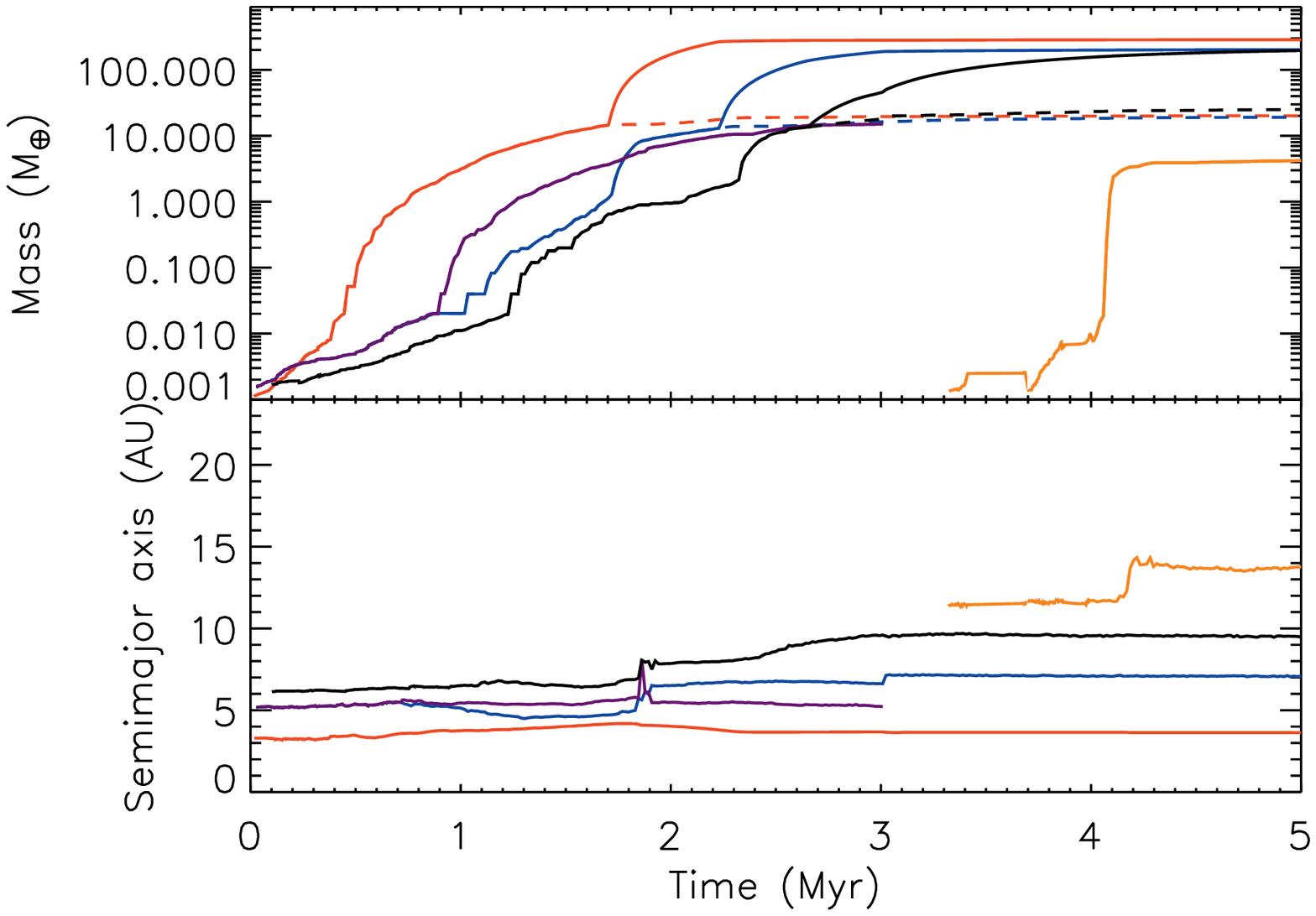}
\caption{The same as Fig.~\ref{fig:run1} but for Run~10.}
\label{fig:run10}
\end{center}
\end{figure}

For Run~10, we adopt large initial planetesimals ($M_0 = 10^{25}$ g) 
and the planetesimal formation timescale 10 times longer ($\tau_{\rm plan1} = 1000$ yr) than that for Run~1.
In Run~10,  pebble accretion is efficient immediately after formation of planetesimals.
In the end of the run,   three gas giant planets form.
Since the planetesimal formation timescale is long, many pebbles are not converted into planetesimals during radial drift and end up  
accumulating at outer edges of gaps opened by massive embryos. 
Pebbles accumulating at gap edges are eventually converted into planetesimals.
Although  these planetesimal are mostly ejected from the system 
due to strong gravitational perturbations of nearby embryos, a massive embryo (with a mass of $4M_{\oplus}$) 
forms at the gap edge of the outer most gas capturing embryo at $\sim$ 4 Myr.
We performed additional simulations using the same parameters if 
two more more ice giant planets form at the gap edge so that they may result in Uranus and Neptune analogs.
We found that formation two or more massive embryos at the same gap edge can occur but 
they end up merging into a single embryo in most cases. 

Growth of embryos' cores after they start runaway gas accretion is much less than that in Run1, since
the population of planetesimals is much lower in Run~10 than that in Run~1.
The mass distribution of remaining planetesimals at $>$ 10 AU is almost the same as the original size distribution, like Run~1.

\section{Parameter dependence}

\begin{figure}
\begin{center}
\includegraphics[width=0.8\textwidth]{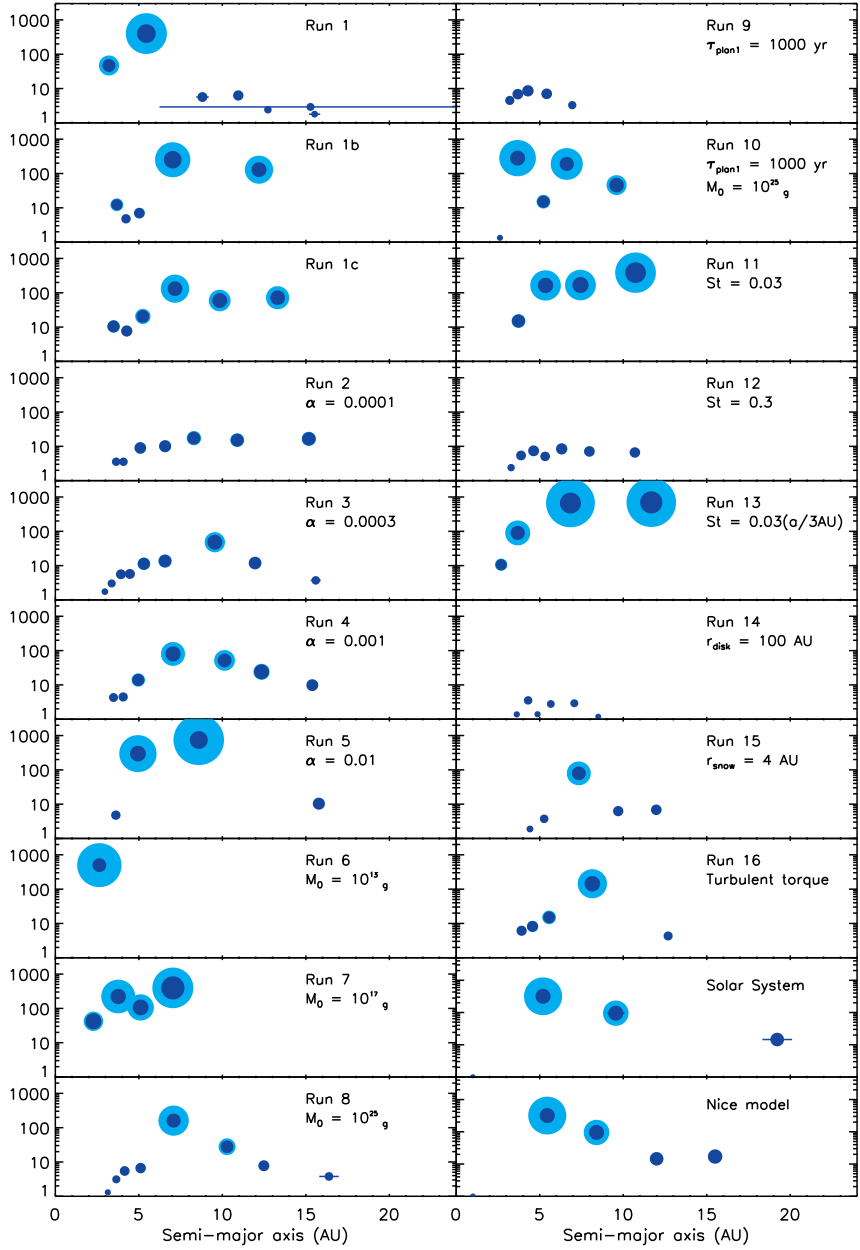}
\caption{Orbital distribution of embryos at 3 Myr for all runs. Only massive embryos ($M > M_{\oplus}$) are shown. 
For reference, the Solar System and the initial condition of the Nice model \citep{Tsiganis2005} are also shown.
For Run~2 through Run~16, the parameters values that are different from Run~1 are shown on the top right of each panel.
The horizontal bar for each body shows orbital excursion due to a finite orbital eccentricity.
}
\label{fig:runall}
\end{center}
\end{figure}

\begin{figure}
\begin{center}
\includegraphics[width=0.7\textwidth]{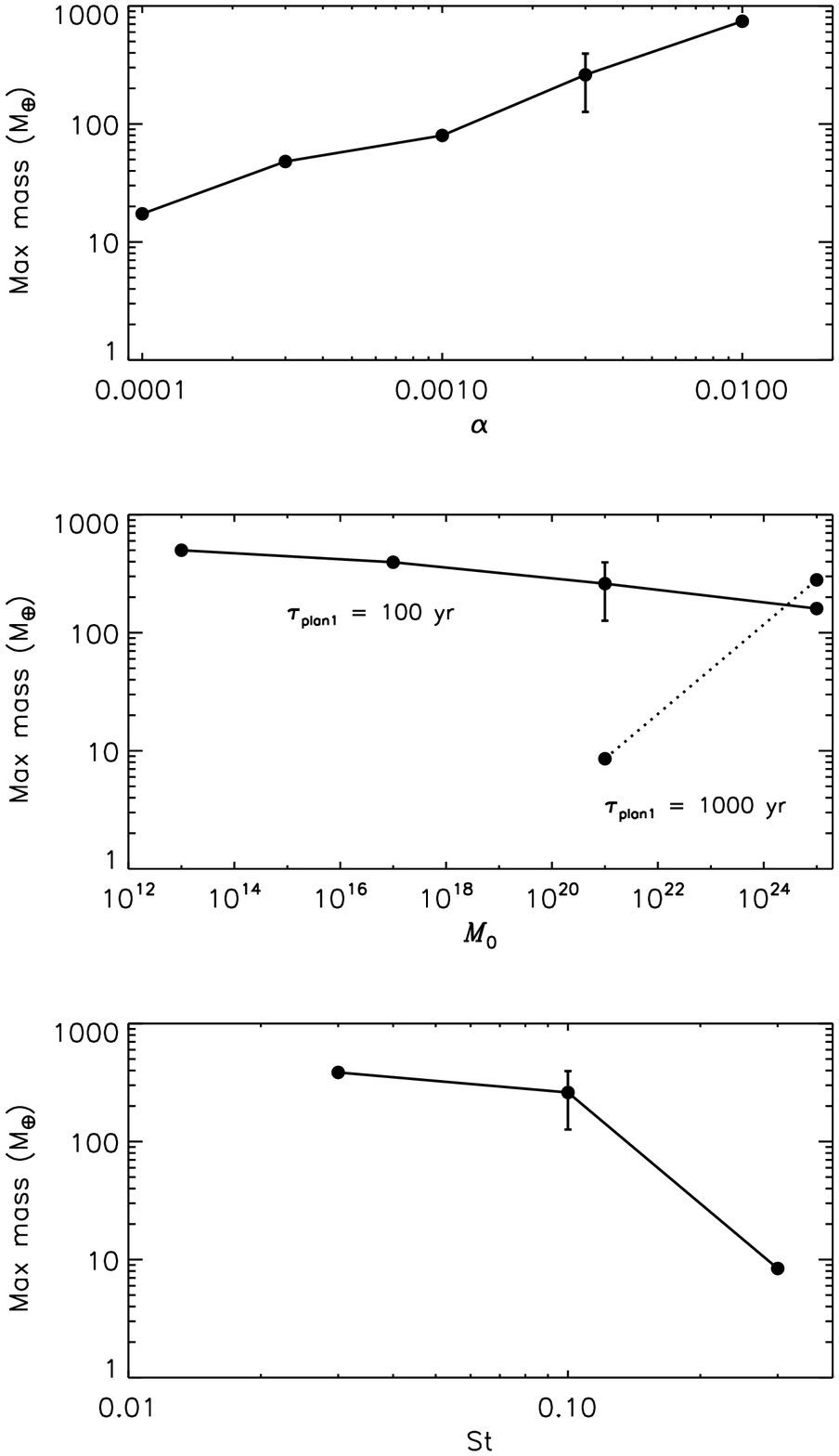}
\caption{The mass of the most massive embryo at 3 Myr as a function of 
(top) $\alpha$, (middle) $M_0$, and (bottom) St.
For the case with the base-line input parameters, we use the mean value and the standard deviation for Runs~1, 1b, and 1c.}
\label{fig:mmax}
\end{center}
\end{figure}

In this section, we discuss how architectures of planetary systems depend on input parameters. 
Figure~\ref{fig:runall} shows architectures of planetary systems at 3 Myr for all the runs.
Figure~\ref{fig:mmax} shows the mass of the largest embryo 
as a function of key input parameters, $\alpha$, $M_0$, and St.

\subsection{Reproducibility}
We first discuss reproducibility for runs with the same input parameters.
Since the statistical routine of our simulation code uses  random numbers in several places, 
results cannot be the same even with the same initial conditions and the same input parameters.
Run~1b and Run~1c have the same input  parameters as Run~1.
The numbers of gas giant planets and 
the masses of the largest gas giant planets are similar to each other for all the three runs.
On the other hand, locations of small embryos are different; In Run~1b and Run~1c, there
are no ice giant analogues outside the outermost gas giant planet, unlike Run~1 and our solar system.  

This difference is found to be caused by 
stochasticity of planetesimal-driven migration of massive embryos. 
In Run~1, the largest embryo forms first near the snow line and embryos of ice giant analogues form later at slightly outer regions ($>$ 5 AU).
In Run~1b and Run~1c,  on the other hand, a large embryo migrates outward and stays around 7 AU, 
stirring planetesimals there and preventing later formation of embryos there.


\subsection{Turbulent viscosity}
The effect of the turbulent viscosity has been already discussed in Section~4.2. 
As $\alpha$ increases, the mass of the largest embryo increases (the top panel of Fig.~\ref{fig:mmax}). 
This is because we adopt a gas surface density independent of $\alpha$ (Eq.~(\ref{eq:sigg}))
and the global gas accretion rate is simply proportional to $\alpha$.
The locations of the largest  embryos are generally at 7 -- 10 AU regardless of $\alpha$.  
The timescale of planetary accretion is too long for formation of massive gas giant planets beyond $\sim$ 10 AU. 

\subsection{Initial planetesimal mass}

Run~6 adopts very small initial planetesimals, $M_0 = 10^{13}$ g.
This run has only one planet inside the snow line in the end. 
Pebbles at the gap edge push the gap opening embryo toward the central star.
In most of runs, however, their total mass is usually much less than the embryo's mass,
as pebbles are converted into planetesimals and they are subsequently ejected by the embryo.
If planetesimals are small enough like the case of Run~6, they cannot 
be ejected as gas drag on small planetesimals are strong.
Small planetesimals are quickly pulverized through collisional destruction to pebble-size fragments.
Thus, the total amount of pebbles trapped at the gap edge increases with time and start to push the embryo very strongly. 
Since we assume that all pebbles sublimate at the snow line, the inward migration of the embryo stops at slightly 
inside the snow line.

For $\tau_{\rm plan1} = 100$ yr, the mass of the most massive embryo decreases with increasing $M_0$
but only weakly (the middle panel of Fig.~\ref{fig:mmax}). 
For $\tau_{\rm plan1} = 1000$ yr, gas giant planets form only for the case of $M_0 = 10^{25}$ g. 
Initially small planetesimals ($M_0 \le 10^{21}$ g) grow very slowly if the population of planetesimals is small.
For $M_0 = 10^{25}$ g, the maximum mass of the largest embryo is larger for longer $\tau_{\rm plan1}$.
This is because the less frequently pebbles are converted into planetesimals, the larger the pebble flux to the largest embryo.

\subsection{Pebble Stokes number (or size)}
The mass of the largest embryo decreases with increasing St (the bottom panel of Fig.~\ref{fig:mmax}).
For the case of St = 0.3, only several ice giant planets form without any gas giant planet.
The dependence  of the largest body's mass on St is caused by two reasons. 
First, efficient pebble accretion can occur for smaller planetesimals for smaller St.
The transition from the hyperbolic regime to the settlement regime occurs if
St $<$ St$^{\ast} \propto M$ \citep{Ormel2010c}, where St$^{\ast}$ is the critical St number.  
Second, the pebble accretion timescale for massive embryos is shorter for smaller St. 
For the 2D case ($R_{\rm coll} \ge h_{\rm pe}$), the pebble accretion rate $\dot{M}$ 
is proportional to St$^{-1/3}$ for a given pebble flux \citep{Morbidelli2015}, while it is proportional to St$^{1/2}$
for the 3D case ($R_{\rm coll} < h_{\rm pe}$). 
Since the growth timescale is longer for massive embryos  in the 2D regime
than for small embryos in 3D regime (see Fig.~\ref{fig:f_coll}), 
the overall timescale for a 1000 km-size planetesimal to grow to a massive embryo 
through both the regimes is shorter for smaller St. 


We adopt constant pebble size for Run~13 instead of constant St for all other runs.
The value of St at the snow line for Run~13 is 0.03, or the same as that for Run~11.
We find  similar masses of the largest bodies for these two runs.
This implies that an important parameter is St around the snow line rather than the radial dependence of St.

\subsection{Other parameters}

The disk size is 100 AU for Run~14 while it is 200 AU for all other runs.
The most massive planet produced in Run~14 is only $\sim 3 M_{\oplus}$, since the pebble flux
quickly ceases for small disks.
Whether a protoplanetary disk is large or not is highly likely one of the most essential criteria 
for formation of gas giant planets. 

The location of the snow line is 4 AU for Run~15 while it is 3 AU for all other runs.
The mass of the largest embryo is smaller in Run~15 than Run~1.
This is because the growth timescale of planetesimals due to mutual collisions increases with 
distance. If $M_0 = 10^{25}$ g, the effect of snow line location is probably relatively less important 
since the radial dependence of growth timescale by pebble accretion is relatively weaker than that by mutual collisions.

As shown by the snapshot of Run~16, the mass of the largest embryos becomes smaller with 
turbulent torques than that without the torques. Turbulent torques excite velocity dispersion of planetesimals 
and slow down their growth.
The effect of turbulent torques is insignificant for massive planetesimals or embryos.

\section{Discussion and future work}
One of our findings is that 
gas giant planets form in disks with a high viscous parameter $\alpha$
while systems with super-Earth- to Neptune-size planets tend to 
form in disks with a low $\alpha$.
On the other hand, \citet{Chambers2016} found that gas giant planets
tend to form in disks with a low $\alpha$ and that 
super-Earth-size planets do not form in his simulations, except in the presence of gas giant planets.
The primal reason causing the different outcomes 
is different types of disk models applied in 
our and his simulations.
While we adopted the same surface density profile independent  
of $\alpha$ provided that the disk surface density evolves in a non-viscous fashion,  
\citet{Chambers2016} adopted purely viscously evolving disks using 
the analytic prescription of \citet{Chambers2009}.
The disk accretion rate is proportional to $\alpha$
in our model while it 
remains moderately high even for a low $\alpha$ 
in a viscously evolving disk due to a high surface density.
\citet{Matsumura2017} adopted the disk accretion rate 
fitted to the observational data (Hartmann et al. 1998) independent  
of $\alpha$ in their simulations. 
Thus, the disk surface density increases with decreasing 
$\alpha$ in their model, and the basic trend of their outcomes is similar 
to that seen in \citet{Chambers2016}.

It is difficult to conclude which disk models are more appropriate than
others at this moment even though our understanding of protoplanetary disks 
rapidly advances, particularly owing to recent disk observations by 
the Atacama Large Millimeter Array (ALMA).
The key parameter to understand angular momentum transport mechanisms
of protoplanetary disks is the relationship between 
the disk mass and the disk accretion rate.
\citet{Rafikov2017} found 
no substantial correlation between 
the disk mass and the disk accretion rate for 26 samples resolved by ALMA \citep{Ansdell2016} and that
$\alpha$ varies from $10^{-4}$ to 0.04.
One of his interpretations is that $\alpha$ is controlled by some yet uncertain mechanisms 
while the disk surface density evolves in a non-viscous manner, somewhat similar to the disk model we assumed.
However, he also suggested other possibilities such as decoupling the global disk accretion rate at large distances from the central star
and the gas accretion rate to the central star. 
Moreover, other authors claim some levels of correlation between 
the disk mass and the disk accretion rate \citep{Manara2016,Lodato2017,Mulders2017}, 
and classical viscous models can be still compatible with observations. 
Since the disk mass is often estimated from the dust abundance assuming the interstellar gas-dust ratio, 
direct measurements of gas densities probably help constrain which disk models are appropriate.
Particularly, measurements of HD abundances \citep{Bergin2013,McClure2016} 
in addition to CO abundance measurements are necessary to 
narrow down uncertainties of observed gas densities.

One of uncertain but intriguing parameters is  the initial mass of the largest planetesimal $M_0$.
For the Solar System, it is probably possible to constrain $M_0$ by 
comparing the size distribution of remaining planetesimals with that for Kuiper belt objects.
The mass distribution per logarithmic size for Kuiper belt has a peak at $\sim$ 100 km in size \citep{Fraser2014,Adams2014}, 
potentially implying that this is the initial size of planetesimals \citep{Morbidelli2009}. 
However, the mass distribution for Kuiper belt has another peak at the Pluto-size, $\sim 3000$ km, 
as the Pluto-size objects contain 10-50 $\%$ mass of the Kuiper belt \citep{Nesvorny2017}. 
If $M_0 = 10^{21}$ g (100 km-size planetesimals), pebble accretion produces a very steep size distribution above 100 km \citep{Johansen2015a}, 
or too small a mass fraction of Pluto-size objects, although the steep slope is consistent with the slope for 
Kuiper belt objects between 100 km and $\sim 300$ km.
Similar results are seen in most of our runs with $M_0 = 10^{21}$ g. 
Run~2, in which we adopted a very low turbulent viscosity, 
exceptionally showed a non-negligible population ($\sim 5\%$) of the Pluto-size objects, which form
at the edges of gaps opened by massive embryos.
Unfortunately, massive embryos quickly form at gap edges and their gravitational perturbations make 
orbits of Pluto-size objects highly eccentric, unlike nearly circular orbits seen in the initial state of the Nice model \citep{Tsiganis2005}. 
Our study implies that the Pluto-size objects in the Solar System are unlikely to have formed from 100 km-size objects through pebble accretion. 
There might have been two distinctive planetesimal formation mechanisms each of which produces 100 km-size and 3000 km-size 
planetesimals in the outer Solar System.


We assumed in this study that all pebbles are icy so that they quickly sublimate inside the snow line.
It is likely, however, that rocky pebbles exist inside the snow line, as evidenced by 
chondrules in chondritic meteorites.
These pebble are expected to be much smaller than icy pebbles due to inefficient mutual sticking. 
If it is the case,  planetary growth by pebble accretion inside the snow line 
is inefficient due to a large scale height of small pebbles \citep{Morbidelli2015}. 
It is of interest to examine if dichotomy of planetary growth across 
the snow line is seen in simulations using our code.
It is computationally demanding, however, for our code to handle small pebbles, as 
the time step for accurate orbital integration proportionally decreases with decreasing St.

Using molybdenum and tungsten isotope measurements on iron meteorites, 
\citet{Kruijer2017} demonstrated that meteorites were derived from two genetically distinct 
nebular reservoirs spatially separated at  $\sim$ 1 My after Solar System formation. 
They suggested that the separation was possibly caused by gap opening of proto-Jupiter. 
Such early formation of Jupiter might be difficult to reconcile with 
chondrule formation ages ranging over a few Myr \citep{Amelin2002,Connelly2012}, since 
chondrule-size pebbles are expected to quickly depletes inside the gap in the absence of 
pebble supply from the outer part of the disk.
This problem might be resolved if the radial structure of the proto-solar nebula 
was very different from those assumed by standard models including the one adopted in this study.
For example, the proto-solar nebula might have the inner cavity inside the asteroid belt 
so that chondrule-size pebbles piled up at the outer edge of the cavity.





\section{Conclusion}
In the present paper, we performed numerical simulations for formation of planetary systems taking into account pebble accretion. 
We varied the size of a protoplanetary disk, the turbulent viscosity, the pebble size, the planetesimal formation efficiency, 
and the initial mass distribution of planetesimals. 
If the planetesimal size exceeds 1000 km,
pebble accretion becomes dominant for planetary growth than by planetesimal accretion. 
If the initial planetesimal size is 100 km, the formation efficiency of planetesimals from pebbles need to be high to 
quickly form 1000 km-size planetesimals. 
Our simulations suggest that planetary systems like ours form from protoplanetary disks with moderately 
high turbulent viscosities. 
If the disk turbulent viscosity is low enough, a planet opens up a gap in the gaseous disk and halts accretion of pebbles 
even before the onset of runway gas accretion. 
Formation of new planets at the gap edges is very efficient for cases of low turbulent viscosities.
Such a disk produces a planetary system with several Neptune-size planets. 
The size distribution of remaining planetesimals is almost the same as the initial size distribution, except for 
the case of very low $\alpha$.
This low-$\alpha$ case showed formation of a non-negligible population of the Pluto-size objects for
the initial planetesimal size of $\sim$ 100 km, although most of 
the Pluto-size objects are highly eccentric unlike the initial condition suggested by the Nice model.
There might have been two distinctive planetesimal formation mechanisms each of which produces 100 km-size and 3000 km-size 
planetesimals in the Kuiper belt.


\section*{Acknowledgements}
This research was carried out in part at the Jet Propulsion Laboratory, California Institute of Technology, 
under contract with NASA. Government sponsorship acknowledged. 
Simulations were performed using a JPL supercomputer, Aurora.

\section*{Data sets}
The source code is available at https://github.com/rmorishima/PBHYB.

\clearpage

\section*{Appendices}

\subsection*{Appendix A: Gas drag coefficient $C_{\rm D}$}

We adopt the numerical coefficient $C_{\rm D}$ of \citet{Adachi1976} 
that covers all the ranges of the Mach number $M_{\rm c} $ and the Knudsen number $K_{\rm n}$.
The Mac and Knudsen numbers $M_{\rm c} $ and $K_{\rm n}$ are given as
\begin{equation}
M_{\rm c} = \frac{v_{\rm rel}}{\gamma^{1/2}c}, \hspace{0.3em} K_{\rm n} = \frac{\ell}{R},
\end{equation}
where $v_{\rm rel}$ is the relative velocity between the particle and ambient gas, $\gamma = 1.4$ is the heat capacity ratio,
$c$ is the isothermal sound speed, $\ell$ is the mean free path of gas molecule, and $R$ is the particle radius.
Using $M_{\rm c}$ and $K_{\rm n}$, the Reynolds number, Re, is given as 
\begin{equation}
{\rm Re} = \frac{6Rv_{\rm rel}}{c_{\rm m} \ell} = \left(\frac{9\pi\gamma}{2}\right)^{1/2} \frac{K_{\rm n}}{M_{\rm c}},
\end{equation}
where $c_{\rm m} (= (8/\pi)^{1/2}c)$ is the thermal velocity of gas.

For $M_{\rm c} \ll 1$ and $K_{\rm n} < 1$, we use the approximated formula given by \citet{Weidenschilling1977} as 
\begin{equation}
C_{\rm D, Wei} = 
\left\{ \begin{array}{ll} 
24 {\rm Re}^{-1}
& \mbox{(for Re $< 1$)}, \\ 
24 {\rm Re}^{-0.6}
& \mbox{(for $1 \le$ Re $< 800$)},\\
0.44 
& \mbox{(for Re $\ge 800$)},\\
\end{array}\right.
\label{eq:cdwei}
\end{equation}
where the case of Re $< 1$ is in the Stokes drag regime.
For $M_{\rm c} \ll 1$ and $K_{\rm n} > 1$, 
the gas drag is in the Epstein drag regime as
\begin{equation}
C_{\rm D, Eps} = \frac{8c_{\rm m}}{3v_{\rm rel}}.
\end{equation}
We give the drag coefficient $C_{\rm D, low}$ for $M_{\rm c} \ll 1$ as
\begin{equation}
C_{\rm D, low} = {\rm Min}(C_{\rm D, Wei},C_{\rm D, Eps}).
\end{equation}
Finally, we interpolate the drag coefficient over $M_{\rm c}$ as \citep{Brasser2007}
\begin{equation}
C_{\rm D} = 
\left\{ \begin{array}{ll} 
C_{\rm D, high}M_{\rm c}^2 +  C_{\rm D, low}(1-M_{\rm c}^2)
& \mbox{(for $M_{\rm c} \le 1$)}, \\ 
C_{\rm D, high}
& \mbox{(for $M_{\rm c} > 1$)},\\
\end{array}\right.
\label{eq:cdi}
\end{equation}
where $C_{\rm D, high} = 2$.

\subsection*{Appendix~B: Turbulent torques}

\citet{Okuzumi2013} derived the stirring rates of planetesimals in turbulence driven by magneto-rotational instability (MRI).
The diffusion coefficient $D_a$ of the semimajor axis $a$ of a body for the ideal MHD case is 
\begin{equation}
D_a = 0.55 \left(\frac{\Sigma_{\rm gas}a^2}{M_{\ast}}\right)^2 \alpha a^2 \Omega,
\end{equation}
where $\Sigma_{\rm gas}$ is the gas surface density, 
$M_{\ast}$ is the mass of the central star, $\alpha$ is the viscosity parameter, and $\Omega$ is the Keplerian frequency at $a$.
At the low eccentricity limit, the change $\delta a$ of $a$ during a time step $\delta t$ is given as 
$\delta a = f_{{\rm den}, \theta} \delta t (2/\Omega)$, where $f_{{\rm den},\theta}$ is the tangential force
due to density fluctuation.
Since $D_a = (1/2) (\langle (\delta a)^2 \rangle/\delta t)$,
we give $f_{{\rm den}, \theta}$  as
\begin{equation}
f_{{\rm den}, \theta} = \frac{\Omega}{2} \xi \left(\frac{2D_a}{\delta t}\right)^{1/2},
\end{equation}
where $\xi$ is the Gaussian white noise with a standard deviation of unity. 
We give the forces in the radial ($r$) and vertical ($z$) directions in a similar manner, 
assuming isotropic turbulence as is the case of aerodynamic turbulent drag.

\subsection*{Appendix~C: Collision probability for the statistical routine}

The expected change rate of the mass $M$ of a planetesimal in the target tracer 
due to merging with planetesimals or pebbles in the interloping tracer $j$ is 
given by \citep{Morishima2015}
\begin{equation}
\left(\frac{dM}{dt}\right)_j= n_j M_j a_{ij}^2 h_{ij}^2  \Omega  P_{\rm col}, \label{eq:dm},
\end{equation}
where $n_j$ is the surface number density of planetesimals/pebbles in the tracer $j$,
$M_j$ is the mass of a constituent planetesimal/pebble, $a_{ij}$ 
is the mean semimajor axis of  the target and the interloper, 
$h_{ij}$ is the reduced mutual Hill radius of the target and interloper, 
and $P_{\rm col}$ is the non-dimensional collision probability (see \citet{Morishima2015} for more details). 
The Hill radius $R_{\rm Hill}$ is given by $a_{ij} h_{ij}$.
In the statistical routine adopted for the study of the present paper,
the target is always a planetesimal-tracer, 
as we do not explicitly handle collisions between pebbles. 
As described in the main text, planetesimals are defined as bodies with Stokes number larger than 2.
Smaller particles are called pebbles.
Different prescriptions are used for pebble and planetesimal interlopers. 

\subsubsection*{C.1. Planetesimal-interlopers}
If the interloper is a planetesimal-tracer (St $>2$) its orbit is well approximated by a Keplerian orbit.
Thus, we calculate the collision probability $P_{\rm col}$ as done in \citet{Morishima2015}. 
We call this probability $P_{\rm col,M15}$ to distinguish from others described below.
This probability corresponds to that for the hyperbolic regime in \citet{Ormel2010c}.

\citet{Ormel2010c} showed that even for St $>2$, due to three body capture, the collision probability 
becomes significantly larger than that for the hyperbolic regime 
if both St and the relative velocity are low enough. 
\citet{Ormel2012a}
derived an empirical form of the collisional radius $b_{\rm 3b}$ normalized by the Hill radius for this effect as 
\begin{equation}
b_{\rm 3b} = \frac{1}{\rm St}\exp{\left[-\left(\frac{0.7\tilde{v_{\rm r}}}{\rm St}\right)^5\right]},
\end{equation}
where $\tilde{v_{\rm r}}$ is the non-dimensional relative velocity.
We give the relative velocity as 
\begin{equation}
\tilde{v_{\rm r}} = \sqrt{\tilde{e}_{ij}^2 + \tilde{i}_{ij}^2},  \hspace{1em} 
\end{equation} 
where $\tilde{e}_{ij}$ and $\tilde{i}_{ij}$ are the relative eccentricity and inclination normalized by $h_{ij}$.
Using $b_{\rm 3b}$, the collision probability $P_{\rm col,3b}$ is given by the smaller of the 3D or the 2D collision probability as 
\begin{equation}
P_{\rm col,3b} = {\rm MIN}(\pi b_{\rm 3b}^2 \tilde{v_{\rm r}}/\tilde{i}_{ij}, 2b_{\rm 3b} \tilde{v_{\rm r}}).
\end{equation} 
The collision probability $P_{\rm col}$ for St $> 2$ is given by 
\begin{equation}
P_{\rm col}({\rm St} > 2) = {\rm MAX}(P_{\rm col,M15},P_{\rm col,3b}).
\end{equation}
If the interloper merges with the target, the velocity of the target is updated in the same manner of \citet{Morishima2015}. 

\subsubsection*{C. 2. Pebble-interlopers}
If the interloper is a pebble-tracer (St $\le$ 2), 
the accretion modes are divided into two regimes depending on the relative velocity. 
The threshold relative speed that separates these two regimes 
is represented by the critical Stokes number ${\rm St}_{\rm crit}$, which is given as \citep{Ormel2010c}
\begin{equation}
{\rm St}_{\rm crit} = {\rm MIN}\left[\frac{12}{\tilde{v}_{\rm r}^3},2\right].
\end{equation}

If ${\rm St} \le {\rm St}_{\rm crit}$ or in the settlement regime,  pebbles settle toward a large body during close encounters with the target.
The non-dimensional collisional radius in this regime is given as \citep{Ormel2012a} 
\begin{equation}
b_{\rm set} = {\rm MIN}\left[\left(\frac{12 {\rm St}}{\tilde{v}_{\rm r}}\right)^{1/2}, 2 {\rm St}^{1/3} \right] \times
\exp{\left[-\left(\frac{\rm St}{\rm St_{\rm crit}}\right)^{0.65}\right]}.
\end{equation}
It is inappropriate to use instantaneous 
Keplerian elements for evaluation of the relative velocity between a pebble-tracer and a target. 
We evaluate the relative velocity in a different manner than that for St $>2$ as follows.
The velocity vector and the radial position of the target are given by $\mbox{\boldmath $v$}$
and $r$, and those for the interloper are given by the same symbols but with the index of $j$. 
For each particle, its velocity relative to the local Keplerian velocity 
$\mbox{\boldmath $v$}_{\rm kep}(r)$ for a body with a circular and non-inclined orbit is calculated as
\begin{equation}
\mbox{\boldmath $v$}' = \mbox{\boldmath $v$} - \mbox{\boldmath $v$}_{\rm kep}(r), \hspace{1em}
\end{equation}
\begin{equation}
\mbox{\boldmath $v$}_j' = \left(\mbox{\boldmath $v$}_j - \mbox{\boldmath $v$}_{\rm kep}(r_j)\right) \left(\frac{r_j}{r}\right)^{1/2},
\end{equation}
where the factor $(r_j/r)^{1/2}$ adjusts the local velocity difference at $r$ and $r_j$.
Using the relative velocity components in the cylindrical coordinates, the non-dimensional relative speed between the target and the interloper is the given as
\begin{equation}
\tilde{v}_{\rm r} = \frac{\left[(v_{r}' - v_{j,r}')^2 + (v_{\theta}' - v_{j,\theta}')^2  + (v_{z}' - v_{j,z}')^2 \right]^{1/2}}{a_{ij}h_{ij}\Omega}.
\end{equation}

If ${\rm St} \ge {\rm St}_{\rm crit}$ or in the hyperbolic regime, the non-dimensional collisional radius is given as
\begin{equation}
b_{\rm hyp} = \tilde{r}_{\rm p}\sqrt{1+\frac{6}{\tilde{r}_{\rm p}\tilde{v}_{\rm r}^2}},
\end{equation}
where $\tilde{r}_{\rm p} = (s+s_j)/(a_{ij}h_{ij})$ is the non-dimensional physical radius 
with the physical radius of a constituent body in the target $s$ and that in the interloper $s_j$. 

The non-dimensional collisional radius for St $\le 2$ is
given by the larger of those in the settlement and hyperbolic regimes as 
\begin{equation}
b_{\rm col} ({\rm St} \le 2) = {\rm MAX}\left[b_{\rm set},b_{\rm hyp} \right].
\end{equation}
The collision probability for St $\le 2$ is given as
\begin{equation}
P_{\rm col} ({\rm St} \le 2) = {\rm MIN}\left[2b_{\rm col}\tilde{v}_{\rm r},\frac{\pi b_{\rm col}^2\tilde{v}_{\rm r}}{\sqrt{2\pi}h_{\rm peb}}\exp{\left(-\frac{z_i^2}{2h_{\rm peb}^2}\right)} \right],
\end{equation}
where $h_{\rm peb}$ is the scale height of pebbles (see the caption of Fig.~\ref{fig:hz} for the exact form of $h_{\rm peb}$).

If the interloper is merged with the target, the velocity of the target relative to the local Keplerian velocity is updated by 
summing up the momenta in the cylindrical coordinates as 
\begin{equation}
\mbox{\boldmath $v$}_{i,{\rm new}}' = \frac{\left[(M v_{i,r}'  + \Delta M v_{j,r}'),  (M v_{i,\theta}'  + \Delta M v_{j,\theta}'), (M v_{i,z}'  + \Delta M v_{j,z}')\right]}{M  + \Delta M}, \hspace{1em}
\end{equation}
where $\Delta M$ is the mass increase of the target planetesimal (Eq.~(13) of \citet{Morishima2015}).

\subsection*{Appendix~D: Capture of planetesimals/pebbles in embryo atmospheres}

\citet{Ormel2012a} derived the approximate analytic solution of the structure of a planetary atmosphere,
provided that heat transport is taken place by radiation only. 
The atmosphere density $\rho_{\rm atm}$ relative to the local nebula density $\rho_{\rm gas}$ is defined as $\sigma = \rho_{\rm atm}/\rho_{\rm gas}$.
The non-dimensional density $\sigma$ as a function of distance $x$ from the planetary center 
is given as
\begin{equation}
\cfrac{s_{\rm Bond}}{x} =
\left\{ \begin{array}{ll} 
1 + \cfrac{2W_{\rm neb}(\sigma-1)+\ln{\sigma}}{\gamma}
& \mbox{(for $\sigma < \sigma_{\rm t}$)}, \\ 
\cfrac{s_{\rm Bond}}{x_{\rm t}} + \cfrac{4}{\gamma}(4W_{\rm neb})^{1/3}\left(\sigma^{1/3} - \sigma_{\rm t}^{1/3}\right) 
& \mbox{(for $\sigma \ge \sigma_{\rm t}$)},\\
\end{array}\right.
\label{eq:sigat}
\end{equation}
where $W_{\rm neb}$ is the non-dimensional parameter 
\begin{equation}
W_{\rm neb} = \frac{3\kappa L P_{\rm gas}}{64 \pi \sigma_{\rm SB} GMT_{\rm gas}^4},
\end{equation}
and $L$ is the luminosity of the planetary atmosphere
\begin{equation}
L= \frac{GM \dot{M}}{R}.
\end{equation}
Here $s_{\rm Bond} = GM/c^2$ is the atmospheric Bondi radius, 
$\kappa$ is the atmospheric opacity,  $P_{\rm gas}$ and $T_{\rm gas}$ are the pressure and the temperature of the local nebula, 
and $\sigma_{\rm SB}$ is the Stefan-Boltzmann constant.

The density $\sigma_{\rm t}$  
at the transition between the nearly isothermal outer region and the warmer inner region is given as
\begin{equation}
\sigma_{\rm t} = \frac{0.2}{W_{\rm neb}}, 
\end{equation}
and the transition radius $x_{\rm t}$ is given by inserting  $\sigma_{\rm t}$ into Eq.~(\ref{eq:sigat}). 
The critical atmospheric density for capturing a body with radius $s_j$
is \citep{Inaba2003a}
\begin{equation}
\rho_{\rm c} = \frac{2}{3}\left(\frac{v_{\infty}^2}{2GM}+\frac{1}{R_{\rm H}} \right) s_j \rho_{\rm p},
\end{equation}
where $v_{\infty}$  is the relative velocity at infinity. 
In our code,  $v_{\infty}$ is approximately evaluated when the separation $x$ is $4.5 R_{\rm H}$ 
during orbital integration. 
The capture radius $s_{\rm atm} = x(\rho_{\rm c})$ is given by setting $\sigma = \rho_{\rm c}/\rho_{\rm gas}$
in Eq.~(\ref{eq:sigat}). If $s_{\rm atm} > s$, $s$ is replaced by $s_{\rm atm}$ in the calculation of $P_{\rm col}$. 
We assume that $s_{\rm atm}$ exceeds neither $s_{\rm Bond}$ nor 0.25$R_{\rm H}$ \citep{Lissauer2009}. 

The atmospheric mass is assumed to be much less than the mass of the solid core for the derivation of 
the above analytic atmospheric structure.  
This is not correct if the embryo is massive enough to trigger runaway gas accretion.
Atmospheric models for massive embryos using 
realistic equation states show that the most of atmospheric mass is concentrated near the solid core \citep{Lee2015}. 
Therefore, it is expected that the structure of 
the less massive outer part of the atmosphere is still approximately given by 
the above analytic formulation, if the core mass is replaced by the total mass of the core and atmosphere.  
We adopt this assumption for embryos in runaway gas accretion in the present study.



\subsection*{Appendix~E: Collisional destruction}
Collisional outcome depends on the specific impact energy $Q$ and the specific energy required 
to disperse half of the mass of the target $Q_{\rm D}^{\star}$.
The energy $Q$ is given as
\begin{equation}
Q = \frac{1}{2}\frac{M M_j  v_{\rm imp}^2}{(M +M_j)^2},
\end{equation}
where $M_j$ is the mass of the impactor and $v_{\rm imp}$ is the impact velocity.
We adopt $Q_{\rm D}^{\star}$ modeled by \citet{Benz1999} as
\begin{equation}
Q_{\rm D}^{\star} = Q_0\left(\frac{s}{1 \hspace{0.2em} {\rm cm}}\right)^j +  B\rho_{\rm p}\left(\frac{s}{1 \hspace{0.2em} {\rm cm}}\right)^k,
\end{equation}
where $s$ is the target radius, $Q_0$, $B$, $j$, and $k$ are the fitting parameters. 
We adopt the values for impacts on ice at $v_{\rm imp} = $ 3 km s$^{-1}$ from Table III of \citet{Benz1999}.
 
We adopt the prescription of the size distribution of ejecta yielded from the impact following \citet{Kobayashi2010b}.
The prescription assumes that impact yields a largest remnant body and a number of small fragments.
The total mass of fragments $M_{\rm frag}$ is given as
\begin{equation}
M_{\rm frag}= (M+M_j)\frac{\phi}{1+\phi},
\end{equation}
where $\phi = Q/Q_{\rm D}^{\star}$.
The mass of the largest remnant $M_{\rm lr}$ is given as
\begin{equation}
M_{\rm lr} = M+M_j -M_{\rm frag}.
\end{equation}
The mass distribution of fragments follows $dn/dM \propto M^{-11/6}$ \citep{Dohnanyi1969}
and the mass of the largest fragment is 
\begin{equation}
M_{\rm lf} = \frac{0.2 M_{\rm frag}}{1+\phi}.
\end{equation}

We apply fragmentation or collisional erosion only to tracer-tracer collisions, not to embryos. 
The procedure to handle fragmentation in the code is as follows. 
We first merge the target and the impactor through the procedure described in \citet{Morishima2015} (see also Appendix~C). 
This gives the position and the velocity of the new tracer. 
We then simply replace the mass of a constituent planetesimal following the 
mass distribution described above and adjust the number of planetesimals
in the tracer so the tracer's mass is unchanged. 
If a random number, which uniformly takes between 0 and 1, is lower than 
$M_{\rm lr} /(M+M_j)$, the new planetesimal's mass is assigned to be $M_{\rm lr}$.
Otherwise, the new planetesimal's mass is a fragment's mass $M$ given by
${\rm rn} = (M/M_{\rm lf})^{1/6}$, where ${\rm rn}$ is another random number between 0 and 1. 
If the Stokes number of the assigned fragment is lower than St for pebbles
we set its mass so that its Stokes number is the pebble's St. 

\clearpage

\section*{References}

\bibliographystyle{model2-names}
\bibliography{giant} 

\end{document}